\documentclass[prd,preprint,floatfix,a4paper,showpacs,aps,amsmath,amssymb]{revtex4}
\usepackage{graphicx}

\begin{document}
\title{Detailed comparison of LIGO and Virgo Inspiral Pipelines in Preparation for a Joint Search}
\author{F.~Beauville$^8$, 
M.-A.~Bizouard$^{10}$, 
L.~Blackburn$^3$, 
L.~Bosi$^{11}$, 
L.~Brocco$^{12}$, 
D.~Brown$^{2,7}$,
D.~Buskulic$^8$, 
F.~Cavalier$^{10}$,
S.~Chatterji$^2$, 
N.~Christensen$^{1,13}$, 
A.-C.~Clapson$^{10}$, 
S.~Fairhurst$^7$, 
D.~Grosjean$^8$, 
G.~Guidi$^{9}$, 
P.~Hello$^{10}$, 
S.~Heng$^6$,
M.~Hewitson$^6$,
E.~Katsavounidis$^3$, 
S.~Klimenko$^5$,
M.~Knight$^1$, 
A.~Lazzarini$^2$,
N.~Leroy$^{10}$,
F.~Marion$^8$, 
J.~Markowitz$^3$, 
C.~Melachrinos$^3$, 
B.~Mours$^8$, 
F.~Ricci$^{12}$, 
A.~Vicer\'e$^9$, 
I. Yakushin$^4$, 
M.~Zanolin$^3$ \\[.3cm]
\begin{center}The joint LIGO/Virgo working group\end{center}}
\vspace{.3cm}
\address{$^1$   Carleton College, Northfield MN 55057 USA}
\address{$^2$   LIGO-California Institute of Technology, Pasadena CA 91125 USA}
\address{$^3$   LIGO-Massachusetts Institute of Technology, Cambridge MA 02139 USA}
\address{$^4$   LIGO Livingston Observatory, Livingston LA 70754, USA}  
\address{$^5$   University of Florida - Gainesville FL 32611, USA}
\address{$^6$   University of Glasgow, Glasgow, G12~8QQ, United Kingdom} 
\address{$^7$   University of Wisconsin - Milwaukee, Milwaukee WI 53201 USA}
\address{$^8$   Laboratoire d'Annecy-le-Vieux de Physique des Particules, Chemin de Bellevue, BP 110, 74941 Annecy-le-Vieux Cedex France }
\address{$^9$   INFN Sezione Firenze/Urbino Via G.Sansone 1, I-50019 Sesto Fiorentino; and/or Universit\`a di Firenze, Largo E.Fermi 2, I-50125 Firenze and/or Universit\`a di Urbino, Via S.Chiara 27, I-61029 Urbino Italia}
\address{$^{10}$   Laboratoire de l'Acc\'el\'erateur Lin\'eaire, IN2P3/CNRS-Universit\'e de Paris XI, BP 34, 91898 Orsay Cedex France}
\address{$^{11}$   INFN Sezione di Perugia and/or  Universit\`a di Perugia, Via A. Pascoli, I-06123 Perugia Italia}
\address{$^{12}$   INFN Sezione di Roma  and/or Universit\`a ``La Sapienza",  P.le A. Moro 2, I-00185, Roma Italia}
\address{$^{13}$ European Gravitational Observatory (EGO), Cascina (Pi), Italia}

\email{ligovirgo@gravity.phys.uwm.edu}

\date{\today}

\begin{abstract}
Presented in this paper is a detailed and direct comparison of the LIGO and Virgo binary neutron star detection pipelines. In order to test the search programs, numerous inspiral signals were added to 24 hours of simulated detector data. The efficiencies of the different pipelines were tested, and found to be comparable. Parameter estimation routines were also tested. We demonstrate that there are definite benefits to be had if LIGO and Virgo conduct a joint coincident analysis; these advantages include increased detection efficiency and the providing of source sky location information.
\end{abstract}

\maketitle

\section{Introduction} 
\label{intro}

The Laser Interferometer Gravitational Wave Observatory (LIGO)~\cite{LIGO-DET} detectors have reached their design sensitivity, while Virgo~\cite{Virgo} is quickly approaching its target sensitivity. This achievement will ultimately be rewarded through the observation of gravitational waves. Numerous potential sources exist, producing signals of differing character. The data analysis efforts of LIGO and Virgo are aiming to detect and identify of these signals.  The inspiral of binary compact objects, such as neutron stars or black holes, is one of the most promising sources of gravitational waves. The observation of the coalescence of binary compact objects will expand our knowledge of the astrophysics of compact objects and provide unique tests of general relativity and cosmology~\cite{shapiro}. LIGO has already conducted searches for binary neutron star inspiral signals, and has placed upper limits on source distributions~\cite{LIGO-IN, LIGO-IN2}.

LIGO and Virgo have each developed methods (analysis pipeline software) for finding binary inspiral signals. In order to maximize the statistical probability of detecting gravitational waves it is likely that LIGO and Virgo will collaborate, and a necessary beginning to such a working relationship will be the validation and understanding of each group's detection strategy. In this paper we present the results of a comprehensive study where the LIGO and Virgo inspiral search pipelines are compared side by side. We have also conducted a similar LIGO-Virgo comparison study for a gravitational wave burst search~\cite{LV2B}. The results presented in this paper show that the LIGO and Virgo binary inspiral detection pipelines operate equally well. In addition, we show that by working together there are undeniable benefits in the quest for detecting gravitational waves from binary inspirals, and that more astrophysical information can be extracted from the signals. For example, the results summarized in this paper demonstrate that a detection strategy based on two-detector coincidence is improved considerably with Virgo included, as opposed to just using LIGO data from the Livingston and Hanford observatories.

In order to conduct a study that compares the capability for inspiral detection it was necessary for both groups to ensure that they were looking for the same signals. LIGO and Virgo groups demonstrated to one another that they were, in fact, looking for the exact same binary inspiral signal. This was a non-trivial commencement to the study, as the nature of the inspiral signal is complicated and only an approximation. Only after this signal generation validation study was complete were we able to mutually verify our detection pipelines. The results of this signal validation effort are presented below.

A previous LIGO-Virgo study showed that both groups were equally competent in finding inspiral signals from {\it optimally oriented} sources~\cite{LV1I}. In the present study here the detection validation is accomplished using simulated data containing {\it realistic} source orientation. Specifically, the binary inspiral sources are simulated to come from galaxies M87 and NGC 6744. Due to the effects of the earth's rotation, and the random orientation of the binary orbital plane, the (simulated) signals impinging on the (simulated) detectors was non-optimum. The present study also incorporates the time delays that will be present in the response of a network of detectors to gravitational waves from some particular sky direction. In this way we hoped to conduct a study whereby the response of the network of the LIGO and Virgo detectors was as realistic as possible. For this study we created simulated noise, with the noise spectral density matching the target expectations for LIGO and Virgo; the simulated noise is Gaussian distributed.

The paper is organized as follows. In Section~\ref{1a} we review the results of our previous analysis where the LIGO and Virgo inspiral pipelines were compared to simple optimally oriented signals. In Section~\ref{siggen} we discuss the comparison of the generation of simulated  binary inspiral signals, and ensure that LIGO and Virgo are searching for signals of the exact same form. In Section~\ref{routines} all of the LIGO and Virgo inspiral search pipelines are discussed, and their response to the data in the present study are presented; the two Virgo pipelines are presented in Section~\ref{MBTA} and Section~\ref{Vflat}, and the LIGO pipeline is described in Section~\ref{LIGO}. The results from a LIGO developed Bayesian parameter estimation technique for detected signals are given in Section~\ref{MCMC}. A side by side comparison of the Virgo and LIGO pipeline results are summarized in Section~\ref{comparison}.  The benefits of a combined LIGO and Virgo inspiral search are presented in Section~\ref{benes}; we find that there is an increase in two-way coincident detection probability, and that there is also the means to gather information on the sky location of the source. Concluding remarks, and an outline for the future goals is presented in Section~\ref{Disc}.

\section{Review of initial comparison project}
\label{1a}
LIGO and Virgo recently initiated a comparison of their inspiral search pipeline~\cite{LV1I}. In this study each group tested their binary inspiral pipelines on simulated data. A similar study was also conducted for the burst search pipelines~\cite{LV1B}. For the inspiral study signals were created from {\it optimally oriented} sources, directly above the interferometers, with noise levels comparable to the target sensitivities for LIGO and Virgo. Both LIGO and Virgo created signals in order to confirm that the other collaboration's pipeline was able to detect them.

The simulated LIGO signals were created using two mass pairs, [1.4 $M_\odot$, 1.4 $M_\odot$], and  [1.0 $M_\odot$, 1.0 $M_\odot$], at various distances (20, 25, 30 and 35 Mpc), and then inserted into the noise. The LIGO h(t) strain signal was created with a sampling rate of 16,384 Hz, and a lower frequency cutoff of 40 Hz. A total of 26 signals were spread over 3 hours of data. The synthesized Virgo signals were from a [1.4 $M_\odot$, 1.4 $M_\odot$] mass pair at a distance of 25 Mpc. The h(t) strain signal had a sampling rate of 20,000 kHz, and a low frequency cutoff of 24 Hz; 9 signals were injected into 2.5 hours of data.

The results from the initial study confirmed the ability of each to correctly detect and characterize a binary inspiral signal.  The LIGO and Virgo groups both analyzed the data created by each group. The LIGO and Virgo pipelines were able to detect the same events, and produced comparable parameter estimates for the chirp mass ($m_c = (m_1 m_2)^{3/5} / (m_1+m_2)^{1/5}$), effective distance, and arrival time. This initial study gave confidence to both groups, and also encouraged us to engage in even more comprehensive and challenging tests. 

\section{Signal Generation}
\label{siggen}
The results of our previous study~\cite{LV1I} were encouraging, but the examination of more realistic signals was needed. We decided to create simulated signals that produce a realistic detection scenario for LIGO and Virgo. When discussing the inspiral of binary neutron star pairs, [1.4 $M_\odot$ - 1.4 $M_\odot$] systems define a convenient reference. For our definition of the sensitivity of a detector we use the distance of a [1.4 $M_\odot$ - 1.4 $M_\odot$] binary pair that is in the optimal direction (directly overhead) and orientation; such a system producing a signal to noise ratio (SNR) of 8 defines our inspiral sensitivity metric. In reality, actual sources will impinge on the detectors from directions, orientations, and polarizations that are sub-optimal. This then produces a decrease in the signal amplitude, or equivalently a larger {\it effective distance}.  When one averages over all directions, orientations and polarizations, the average effective distance to sources is 2.3 times greater than the actual distance~\cite{finn-chernoff}. LIGO and Virgo were designed such that at their target sensitivities their inspiral ranges would extend up to 35 Mpc (for an optimally oriented 1.4 $M_\odot$ - 1.4 $M_\odot$  binary inspiral). Since the Virgo cluster of galaxies is within this distance (at 16 Mpc), we decided to have this LIGO-Virgo inspiral study respond to signals from the M87 galaxy, as it is the largest galaxy in the Virgo cluster. In addition, other signals were created to simulate their emission from NGC 6744 at 10 Mpc. The random orientation of the sources, plus the rotation of the Earth, created signals that produced a wide variety of responses from the detectors.

\begin{figure}[tb]
  \begin{center}
    \includegraphics[width=4.6in,angle=0]{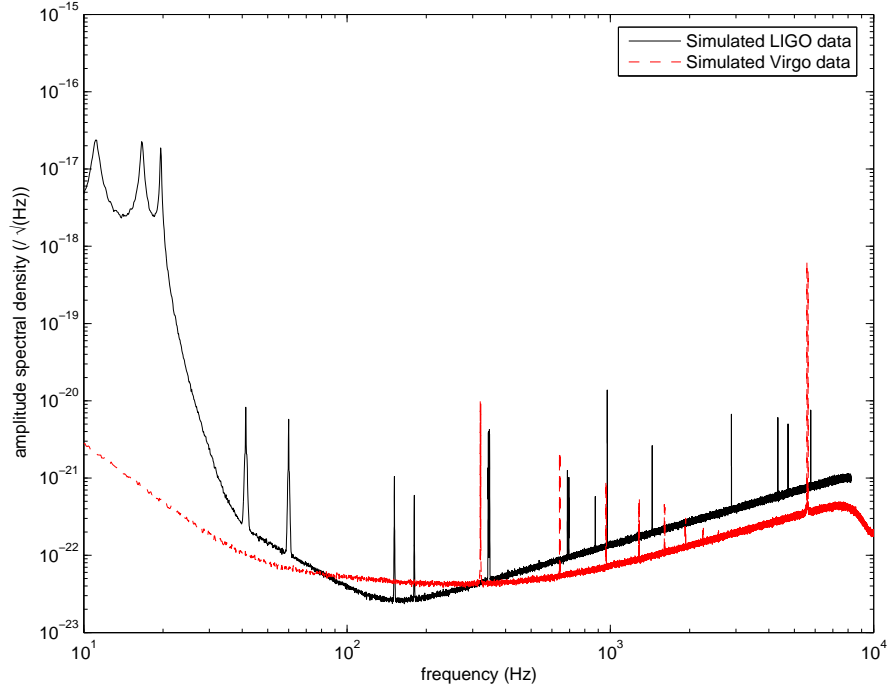}
  \end{center}
\caption{The noise spectrum (at design sensitivity) of the simulated data for LIGO (solid line) and Virgo (dashed line). The distortion close to the Nyquist frequency in the Virgo spectrum is due 
to the use of a low pass filter applied before downsampling the Virgo data generated at 
40 kHz down to 20 kHz. }
\label{fig1}
\end{figure}

For the study presented in this paper the LIGO and Virgo groups each created 24 hours of simulated data. The LIGO detectors modeled were the 4 km systems at Hanford WA (H1), and Livingston LA (L1), while the Virgo detector was that at Cascina, Italy (V1). The noise of the data matched that of the target sensitivities for the LIGO and Virgo interferometers. Fig.~\ref{fig1} shows the noise spectral densities for the simulated data. Within the 24 hours of data 144 inspiral signals were injected from the two galaxies; the sources came from systems with random orbital plane orientations. The masses of the binary system stars were within the $1 M_\odot$ to $3 M_\odot$ range.

In order to initiate our study of the LIGO and Virgo inspiral detection capacities we ensured that the production of 2.0 post-Newtonian (PN)~\cite{blanchet0} binary inspiral signals for 1.4 $M_\odot$ - 1.4 $M_\odot$ mass pairs by both groups was identical. This exercise highlighted three important areas that need to be closely monitored in studies dependent on simulated inspiral signal generation. It is important to use the exact same value of Newton's constant, G. Second, the lengths of the inspiral chirps need to be monitored. Finally, the definition for the termination frequency of the inspiral signal needs to be defined in the same way. Since the signals calculated come from approximation methods in general relativity it is not surprising that slight discrepancies can occur. At 2.0 PN one can write the frequency of the waveform as a function of time $f(t)$, or the time before coalescence as a function of the frequency $t(f)$. These two functions are inverses of each other only to 2.0 PN order. In the LIGO code, the length of the chirp is determined by solving $f(t) = f_{low}$ using a root finder. However, in the Virgo code, the length of the chirp is found by determining $t(f_{low})$. For a 1.4-1.4 solar mass inspiral beginning at 30 Hz, the LIGO code generates a signal that is 54.6789 s, while the inspiral generated by the Virgo software is 54.6799 s, so the calculated length of the chirp differs by 1 ms between the two methods. In the test we are performing, this leads to an offset between the two chirps. It was possible to obtain agreement between the two waveforms by simply sliding the Virgo waveform backwards by 1 ms. For the signal termination there was a slight difference found between the LIGO and Virgo methods, which was due to the two codes using a somewhat different ending frequencies. LIGO uses the test mass innermost stable circular orbit (ISCO). Virgo nominally uses the last stable orbit (LSO, the occurrence of the minimum of the dynamical energy, see~\cite{blanchet0}) calculated at 2PN order, or when the phasing formula defined in~\cite{blanchet} breaks down at 2.0 PN order (which was the case for the Virgo generated signals in present study). This then results in an additional cycle at the very end of the Virgo waveform, but we found that this contributes negligibly to the signal to noise ratio for the binary neutron star coalescence, and the detection results. When one does calculations at a
finite order in a perturbative expansion, it is unavoidable to meet
quantities that do not yield the same value when computed via different
methods, and shifting the Virgo waveform by 1 ms was a way to show that
this perturbative issue can be traded with a redefinition of the time.

The signals generated by LIGO and Virgo in this present study both used the same definition of G, specifically G = $6.67259\times 10^{-11} m^3 kg^{-1} s^{-2}$. The LIGO and Virgo signals were adjusted so that the coalescence times and phases were the same, as these are some of the important parameters to be estimated by the detection pipelines. The sky locations were determined by the galaxies. The angle parameters for the signals (phase at coalescence, polarization, the cosine of the orbital plane inclination) were chosen randomly from uniform distributions. The masses were selected from a small set (1.0, 1.4, 2.0, and 3.0 $M_\odot$). The injections were spaced in time so that they would not significantly affect the power spectrum estimation; the injection times chosen randomly with an average of one injection every 600 s. In the end we were content with the overall similarity of the LIGO and Virgo signals, and proceeded with the detection study.

\section{Virgo and LIGO inspiral detection routines}
\label{routines}
Virgo has developed two search pipelines for binary neutron star inspiral signals, the purpose of which is to experiment with different analysis solutions and cross-check their outputs; the plan
is to keep developing both methods, because it is anticipated that the
two may have different merits when applied to different portions of the
parameter space, and/or to different kinds of binary systems (black holes, inclusion of spin, etc.).
One Virgo method is a multi-band templated analysis (MBTA), whereby the templates are split for efficiency into low and high frequency parts~\cite{MBTAp}. The templates are then subsequently combined together in a hierarchical way. Virgo also has a standard flat-search pipeline, called Merlino~\cite{Merlinop}, that is similar to the LIGO inspiral pipeline~\cite{LIGO-IN,LIGO-IN2}. Both of these Virgo inspiral detection pipelines, as well as the single LIGO pipeline, were applied to the data in this study. 
A summary of the basic details of the Virgo and LIGO inspiral detection pipelines is 
given in Table~\ref{prod_parameters}; detailed descriptions are in the sub-sections below.
The range of masses covered by the templates was 1-3 M$_{\odot}$. The template banks were 
created with a minimal match 
criterion of 95\%, ensuring that no event in that mass space would be 
detected with a SNR loss greater than 5\%.
The starting frequency $f_{low}$ for the analysis of the data, 
in order to be consistent with the SNR accumulation (defined by the sensitivity curves of both 
experiments) was set to 40~Hz for LIGO data and to 30~Hz for 
Virgo data. Triggers were recorded when the SNR exceeded a 
threshold of 6. With the trigger lists from each of the pipelines, an event is labeled as {\it true} if the end time of the inspiral event matches the end time of the injected inspiral event within $\pm 10$ ms.

\begin{table}[h!]
\begin{center}
\begin{tabular}{|l|c|c|}
\hline
 & LIGO dataset & Virgo dataset \\ \hline
 Mass range & 1-3 M$_{\odot}$  & 1-3 M$_{\odot}$ \\ \hline
 Grid minimal match & 95\% & 95\%  \\ \hline
 Starting frequency $f_{low}$ & 40 Hz & 30 Hz \\ \hline
Longest template duration & $\sim$ 45 seconds & $\sim$ 96 seconds \\ \hline
 SNR threshold & 6 & 6 \\ \hline
\end{tabular}
\end{center}
\caption{Common search parameters for the LSC and Virgo pipelines.}
\label{prod_parameters}
\end{table}

\par
The mass parameter space layout for the LIGO grid is described 
in~\cite{Owen_Sathya}. The Virgo MBTA pipeline creates the grid according to a 
2D contour reconstruction technique based on the parameter space 
metric~\cite{Damir}. The Virgo Merlino template placement is 
explained in ~\cite{Merlinop}.
Table~\ref{production} provides information about the way the 
production was done for the three pipelines and each data set, which type of processor was used, and other resources needed.

\begin{table}[h!]
\begin{center}
\hspace*{-1cm}
\begin{tabular}{|l|c|c|c|c|c|c|}
\hline
Pipeline & LIGO & MBTA & Merlino & LIGO & MBTA  & Merlino \\  \hline
Data Set &  LIGO & LIGO & LIGO & Virgo & Virgo & Virgo \\ \hline
Number of templates & $\sim$ 2900 & $\sim$ 1900 & $\sim$ 2000 & $\sim$ 10900 & $\sim$ 7000 & $\sim$ 6800 \\ \hline
Type of processor & 1 GHz & 2.2 GHz & 2.2 GHz & 2.66 GHz & 2.2 GHz & 2.2 GHz \\
& Pentium II & Opteron & Opteron & Xeon & Opteron & Opteron \\ \hline
Total processing time & $\sim$ 368 hours & $\sim$ 55 hours & $\sim$ 50 hours
& $\sim$ 704 hours & $\sim$ 231 hours & 250 hours \\ \hline
$\frac{{\rm processing\ time} \times{\rm processor\ speed}}{{\rm number\ of\ templates}}$
& \raisebox{0cm}[.6cm][.4cm]{$\sim$ 457 s\ GHz} & $\sim$ 229 s\ GHz & 198 s\ GHz & $\sim$ 618 s\ GHz & $\sim$ 261 s\ GHz & 291 s\ GHz \\ \hline
\end{tabular}
\end{center}
\caption{Configuration and computing cost of each analysis.}
\label{production}
\end{table}
 Table~\ref{result1} summarizes the detection efficiency results of the three inspiral pipelines on their application to the data sets used in this study. Table~\ref{result2} provides the information on the accuracy of each of the three pipelines for parameter determination for the signals detected in the V1 data. These results are explained in the sections below.
\begin{table}[h!]
\begin{center}
\begin{tabular}{|l|c|c|c|c|c|c|}
\hline
& \% detected & \% detected &\% detected & FAR for & FAR for \\
 & in L1 data & in H1 data & in V1 data & LIGO data & Virgo data\\ \hline
 LIGO pipeline & 64\% & 66\% & 62\% & 0.07 Hz & 0.77 Hz \\ \hline
 MBTA pipeline & 62\% & 61\% & 56\% & 0.02 Hz & 0.1 Hz \\ \hline
 Merlino pipeline & 55\% & 59\% & 55\% & 0.03 Hz & 0.1 Hz \\ \hline
\end{tabular}
\end{center}
\caption{The signal detection results for the LIGO, MBTA and Merlino inspiral detection pipelines applied to the L1, H1 and V1 data. The false alarm rate (FAR) for the pipelines applied to the data is also listed.}
\label{result1}
\end{table}
\begin{table}[h!]
\begin{center}
\begin{tabular}{|l|c|c|c|c|c|c|}
\hline
& chirp mass & chirp mass & end time & end time & effective distance & effective distance \\
& difference & difference & difference & difference & fractional error & fractional error \\
& mean ($M_{\odot}$) & RMS ($M_{\odot}$) & mean (ms) & RMS (ms) & mean & RMS \\ \hline
 MBTA pipeline &  $2.46 \times 10^{-4} $ & $9.13 \times 10^{-4}$ & 0.340 & 0.876 & -0.0031 & 0.10 \\ \hline
 Merlino pipeline & $1.60 \times 10^{-4}$ & $1.00 \times 10^{-3}$ & 0.083 & 1.03 & 0.01 & 0.11 \\ \hline
 LIGO pipeline & $2.79 \times 10^{-4}$ & $9.04 \times 10^{-4}$ & 0.966 & 1.29 & -0.026 & 0.112 \\ \hline
\end{tabular}
\end{center}
\caption{Parameter determination accuracy results for the LIGO, MBTA and Merlino inspiral detection pipelines applied to V1 data. The values in the table are derived from the distribution of all events detected by the respective pipeline. The parameter difference is defined as the actual parameter value subtracted from the pipeline's recovered value; results are given for the chirp mass (in units of $M_{\odot}$) and end time (in units of s). Also listed is the effective distance fractional error, which is defined as (recovered effective distance - actual effective distance)/(actual effective distance).}
\label{result2}
\end{table}

\subsection{Results for Virgo multi-band analysis of inspiral signals}
\label{MBTA}
The Virgo MBTA inspiral detection pipeline was designed to reduce the computational cost of a binary inspiral search. A large number of templates are needed to sufficiently cover the parameter space, especially for binary systems containing relatively smaller masses. The required number of templates depends on the duration of the longest possible signal, and this is affected by the relatively slow frequency evolution of the binary inspiral at lower frequencies. The computational speed of a Fast Fourier Transform (FFT) depends on the frequency span of the data set. The matched filtering technique uses FFTs as part of the computation. The goal of the MBTA technique is to split the analysis into low and high frequency parts; these results are subsequently combined in a hierarchical way. A description of the method can be found in~\cite{MBTAp}. 

The MBTA pipeline was applied to data from H1, L1 and V1. Detected events with $SNR>6$ were recorded, and clustered both in time and over the template bank; triggers with matching endtimes (within $\pm 10$ ms) were considered the same event, and the trigger with the highest SNR was recorded; if this trigger was within $\pm 10$ ms of an injection event end time then the it was specified as a detection.  The template bank spanned the range from $1 M_\odot$ to $3 M_\odot$ and had a minimal match of 0.95; a total of 7000 templates were used to analyze the Virgo data set, and 1900 templates for the LIGO data. The Virgo MBTA code was run with a splitting frequency 
between the low and high frequency bands chosen so as to share in an 
approximately equal way the SNR between the two bands; the band splitting frequency was 130 Hz when applied to the LIGO data, and 95.3 Hz for the Virgo data. With these settings the single instrument false alarm rate was 0.1 Hz for the Virgo data, and 0.02 Hz for the LIGO data. In general, the higher false alarm rate with the Virgo data (seen with all the pipelines) is due to the larger number of templates used, which itself is due to the lower frequency cut-off of the search. 
Based on the number of signals found in the data set, the single interferometer detection efficiencies for the MBTA pipeline was 61\% for signals in the H1 data, 62\% for events detected in the L1 data, and 56\% for signals in the V1 data. The actual value of the efficiency depends entirely on the source population (and resultant effective distance distribution) chosen. Because we have chosen a population for which the efficiency is roughly 50\%, we are particularly sensitive to small differences in the pipelines and algorithms, which is the goal of this study. There were a total of 144 injections, so the statistical error on the detection efficiencies is around 4\% for each detection pipeline. The injections that were not found had effective distances exceeding the range of the instrument, typically greater than 50 Mpc.

\begin{figure}[tb]
  \begin{center}
    \includegraphics[width=2.6in,angle=0]{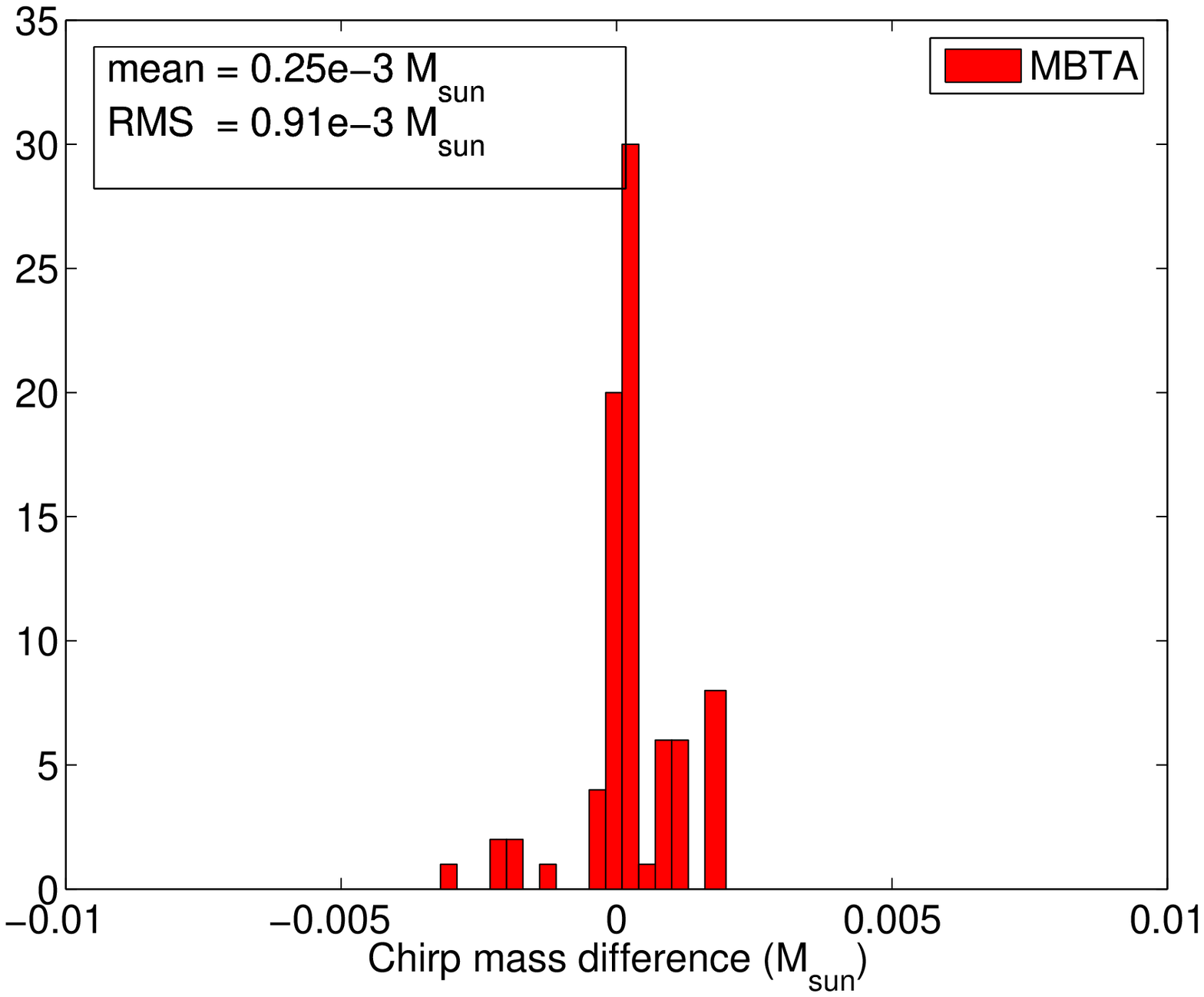}
    \includegraphics[width=2.6in,angle=0]{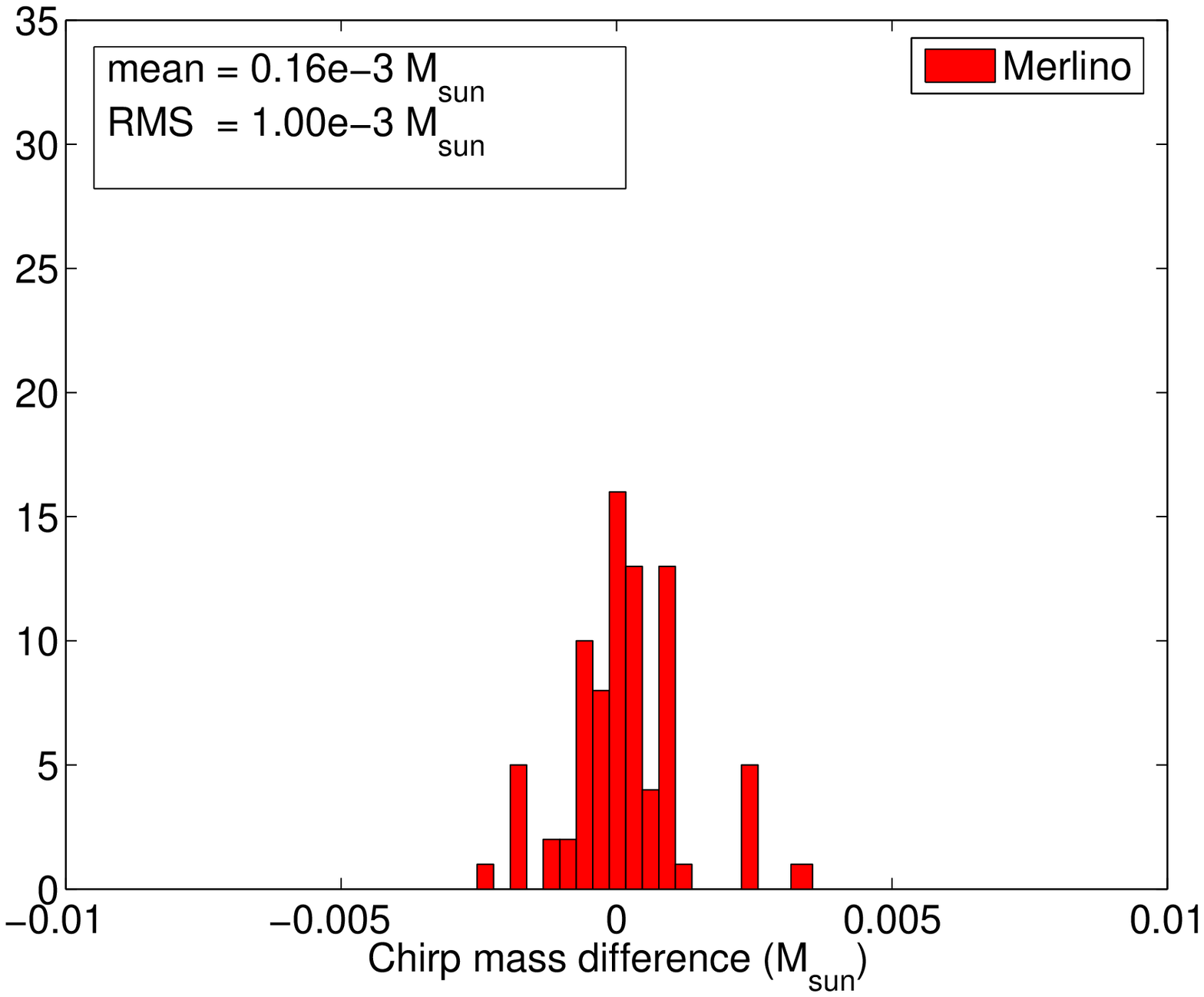}
    \includegraphics[width=2.6in,angle=0]{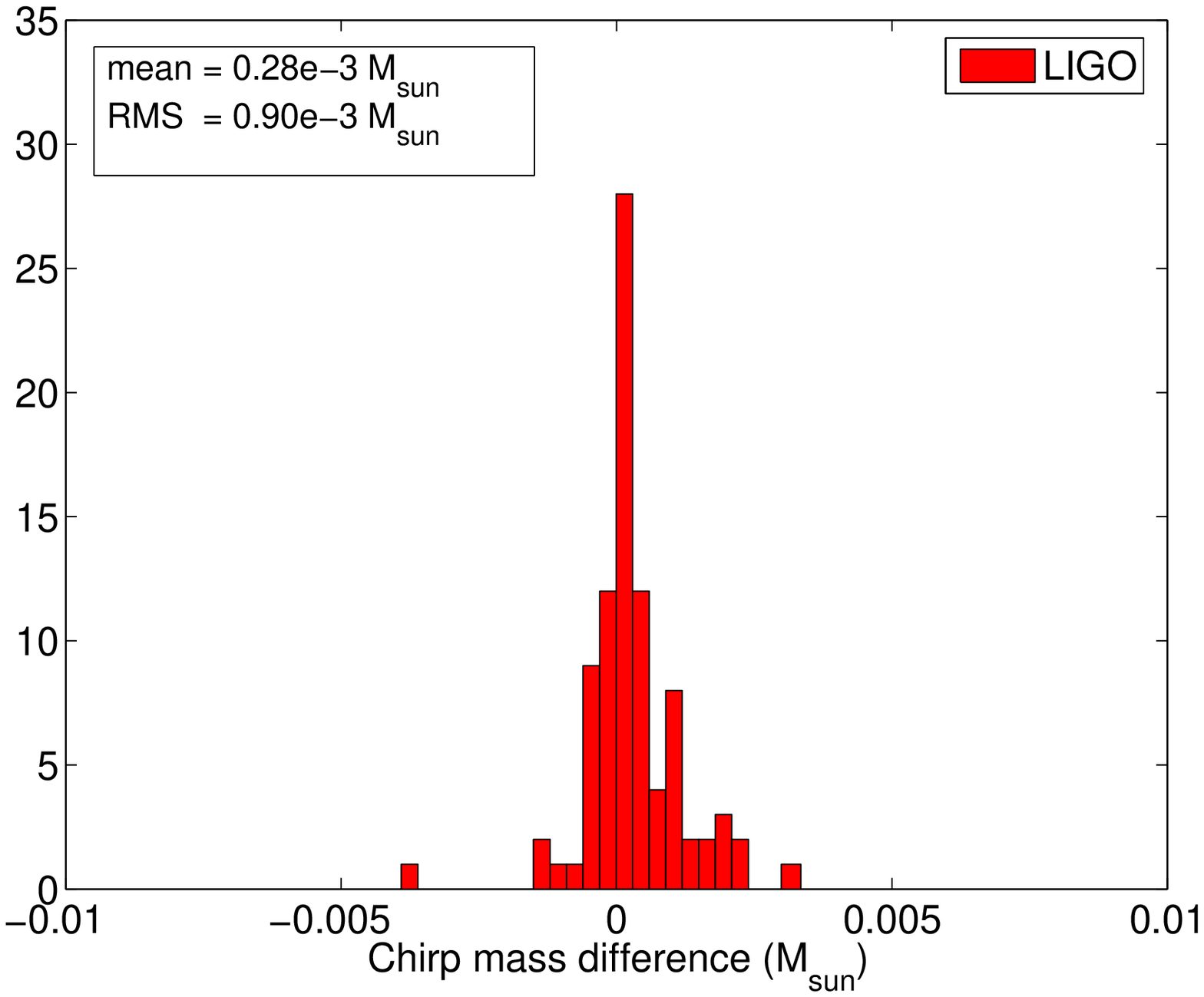}
  \end{center}
\caption{Histograms of the chirp mass determination accuracy for the Virgo MBTA, Virgo Merlino, and LIGO inspiral analysis pipelines. These three pipelines were applied to the Virgo data, and displayed are the results from the signals recovered. The {\it difference} is defined as the actual chirp mass subtracted from the recovered chirp mass.}
\label{Vchirp}
\end{figure}

\begin{figure}[tb]
  \begin{center}
    \includegraphics[width=2.6in,angle=0]{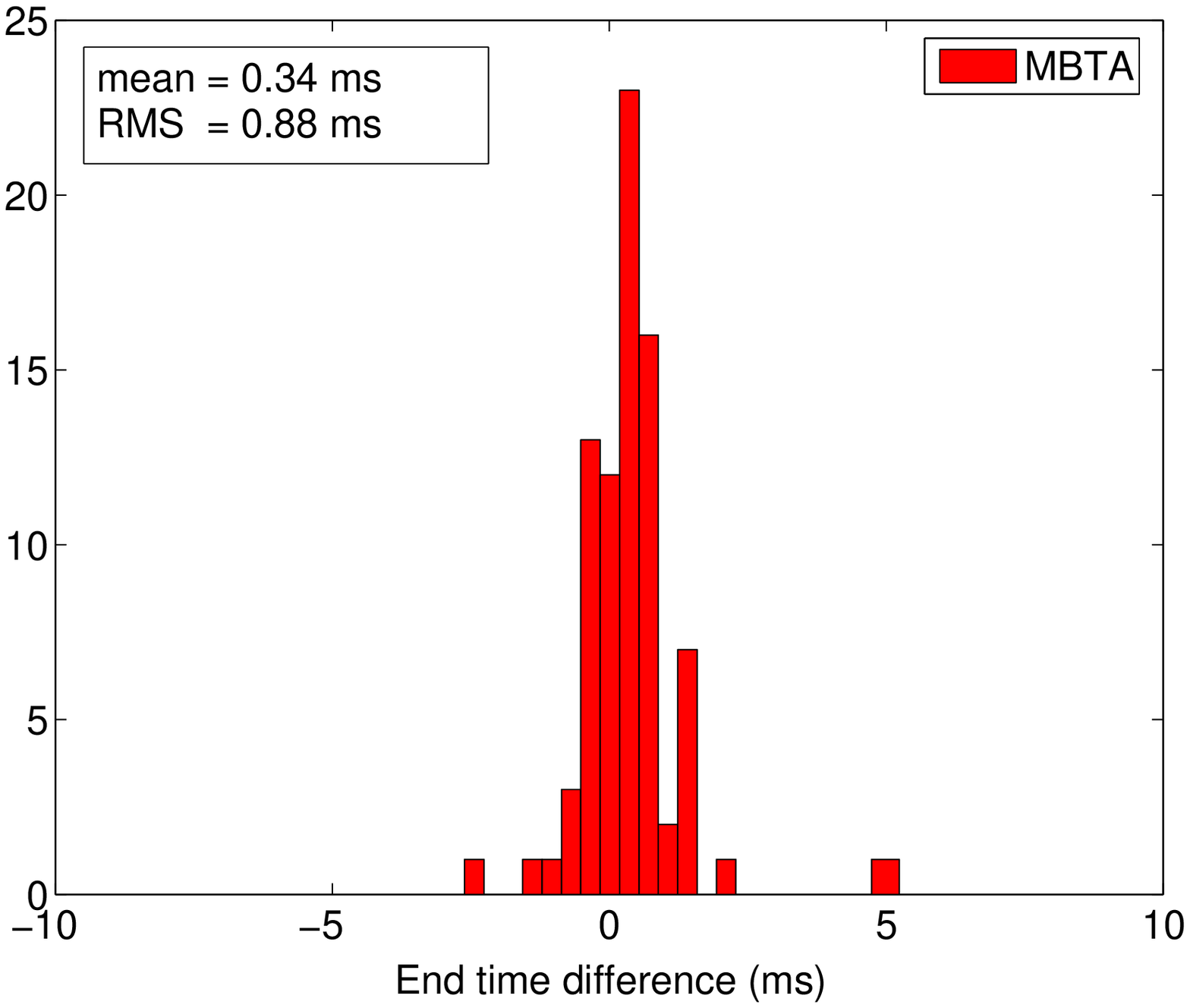}
    \includegraphics[width=2.6in,angle=0]{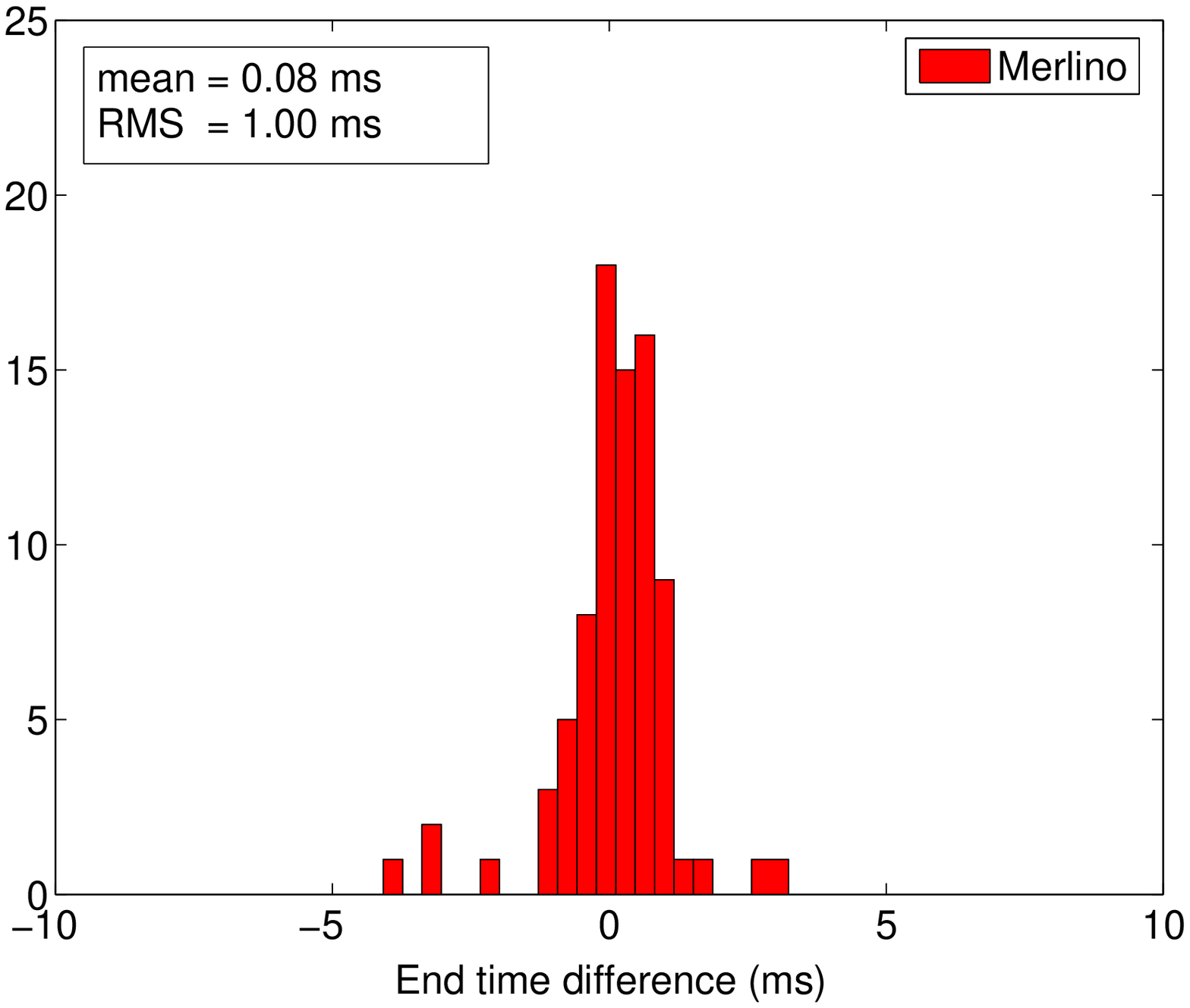}
    \includegraphics[width=2.6in,angle=0]{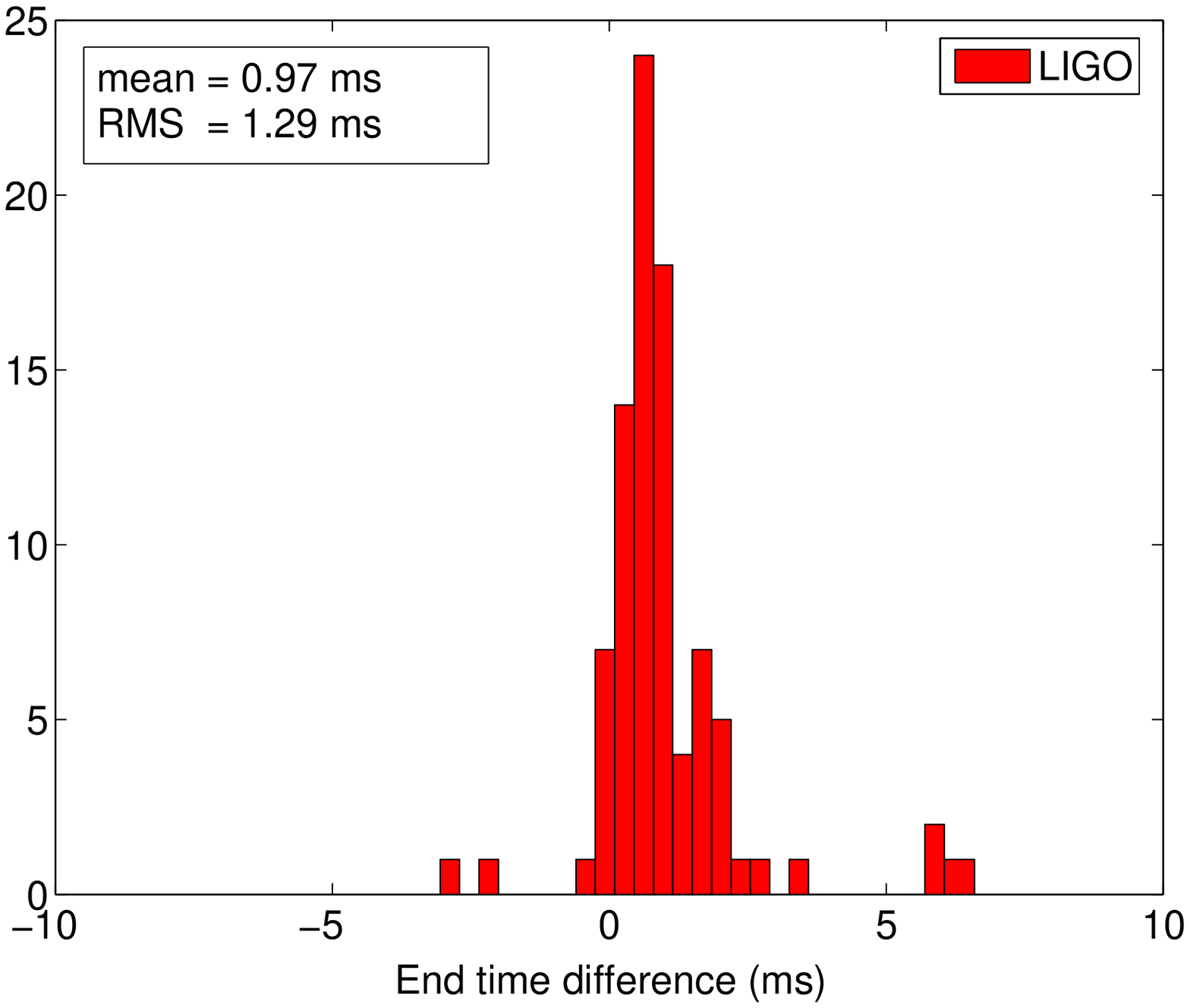}
  \end{center}
\caption{Histograms of the end time determination accuracy for the Virgo MBTA, Virgo Merlino, and LIGO inspiral analysis pipelines. These three pipelines were applied to the Virgo data, and displayed are the results from the signals recovered. The {\it difference} is defined as the actual end time subtracted from the recovered end time.}
\label{Vend}
\end{figure}

\begin{figure}[tb]
  \begin{center}
    \includegraphics[width=2.6in,angle=0]{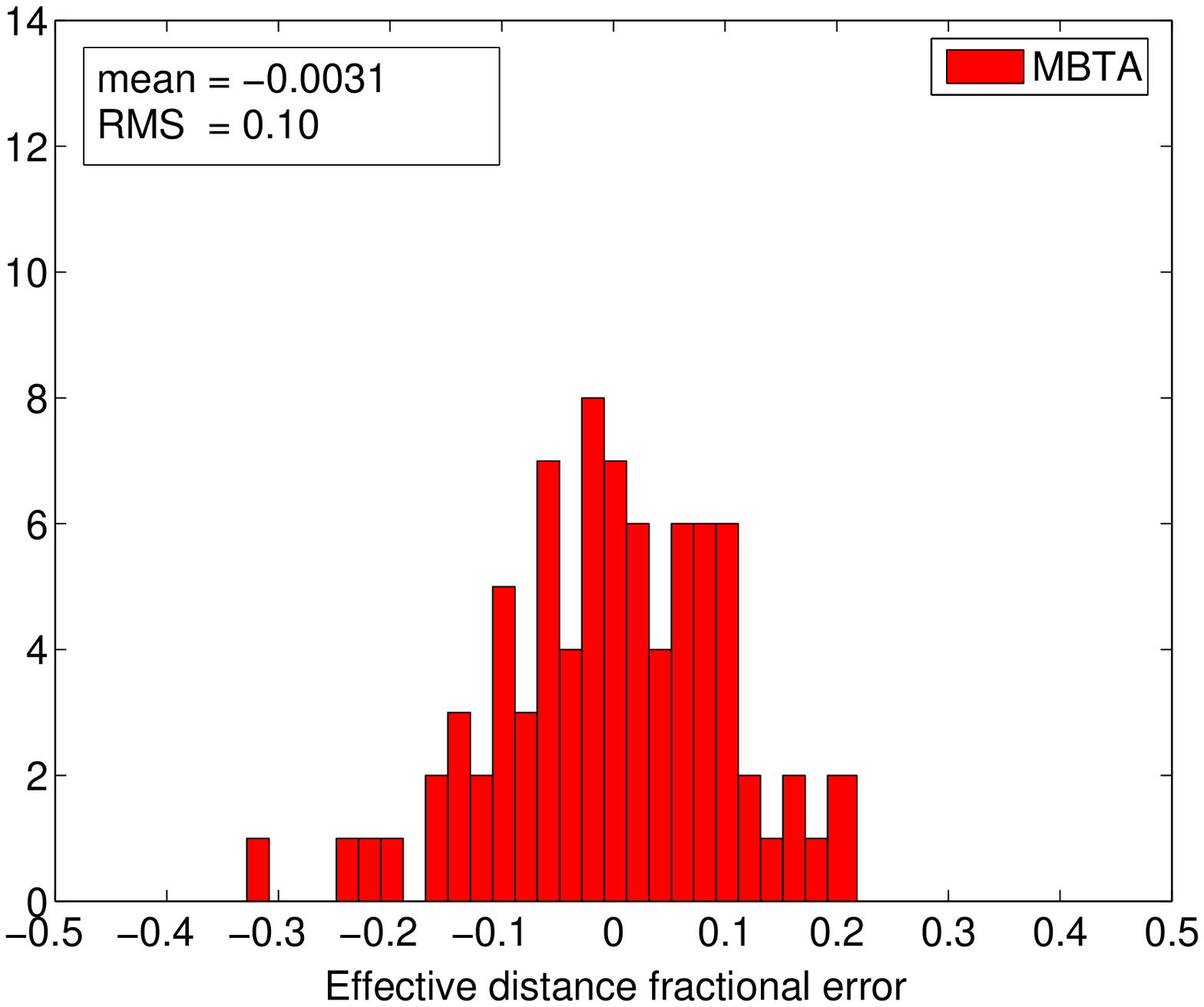}
    \includegraphics[width=2.6in,angle=0]{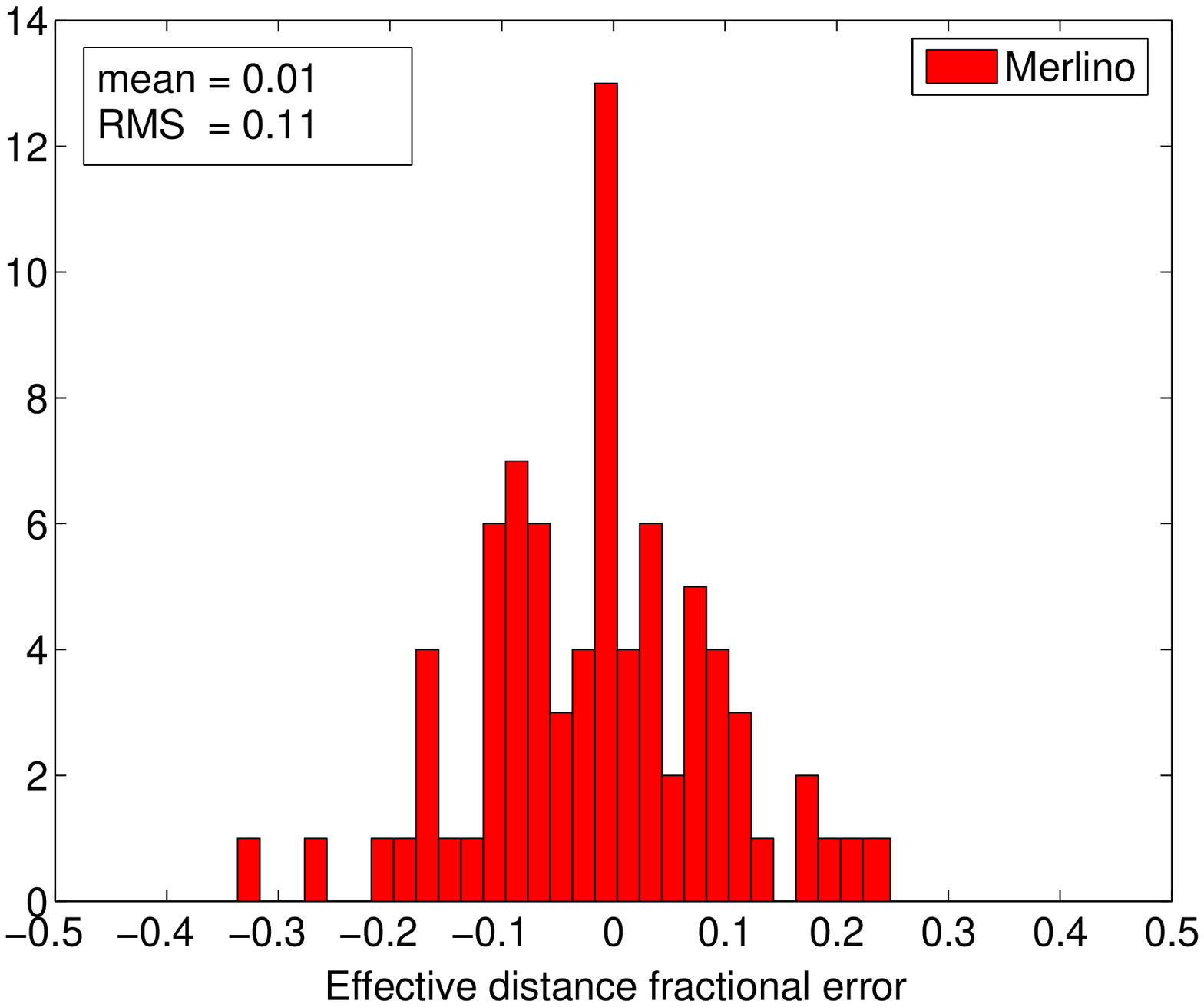}
    \includegraphics[width=2.6in,angle=0]{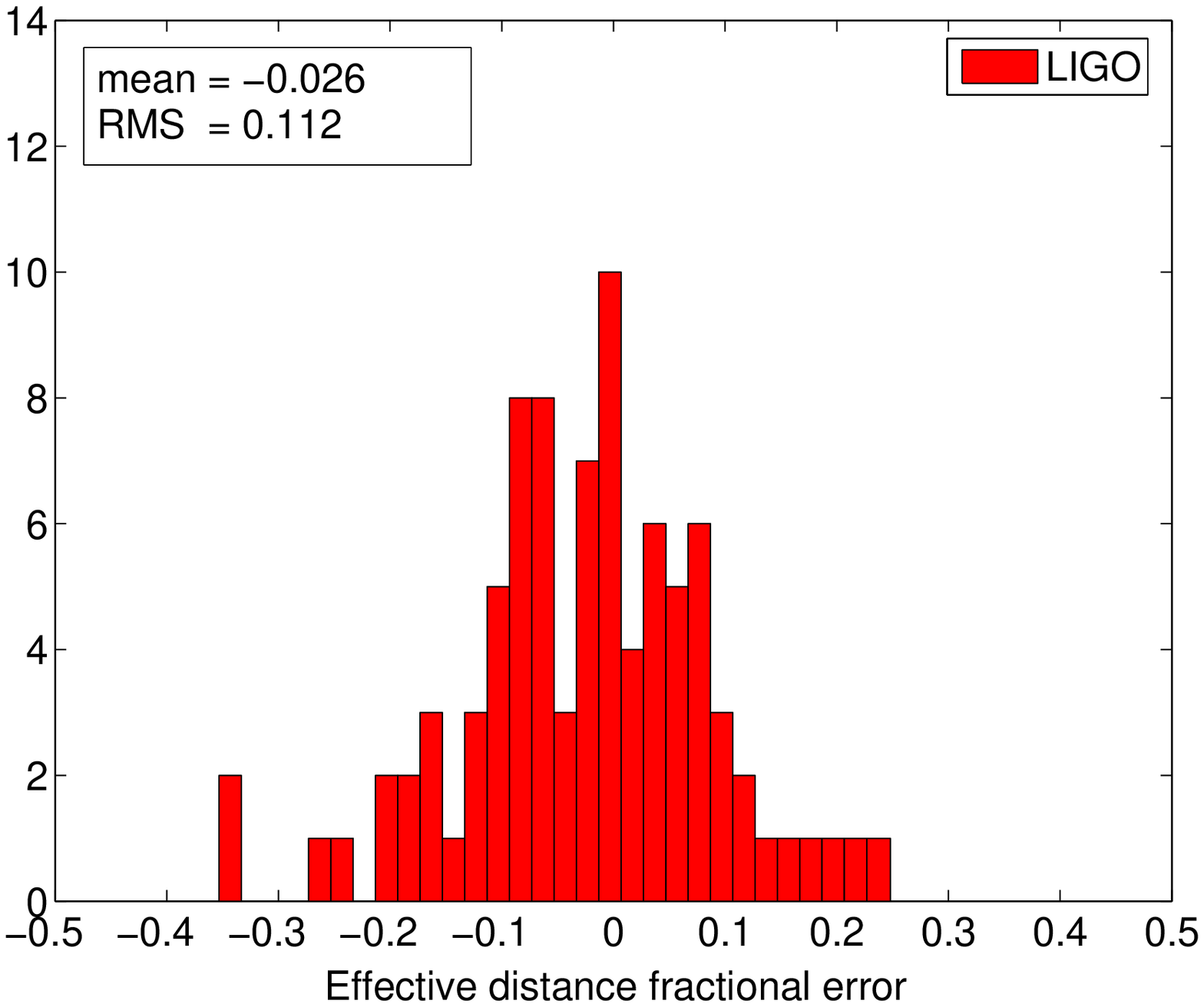}
  \end{center}
\caption{Histogram of the effective distance fractional error for the Virgo MBTA, Virgo Merlino, and LIGO inspiral analysis pipelines. These three pipelines were applied to the Virgo data, and displayed are the results from the signals recovered. The effective distance fractional error is defined as (recovered effective distance - actual effective distance)/(actual effective distance).}
\label{Vdist}
\end{figure}

\begin{figure}[tb]
  \begin{center}
    \includegraphics[width=2.6in,angle=0]{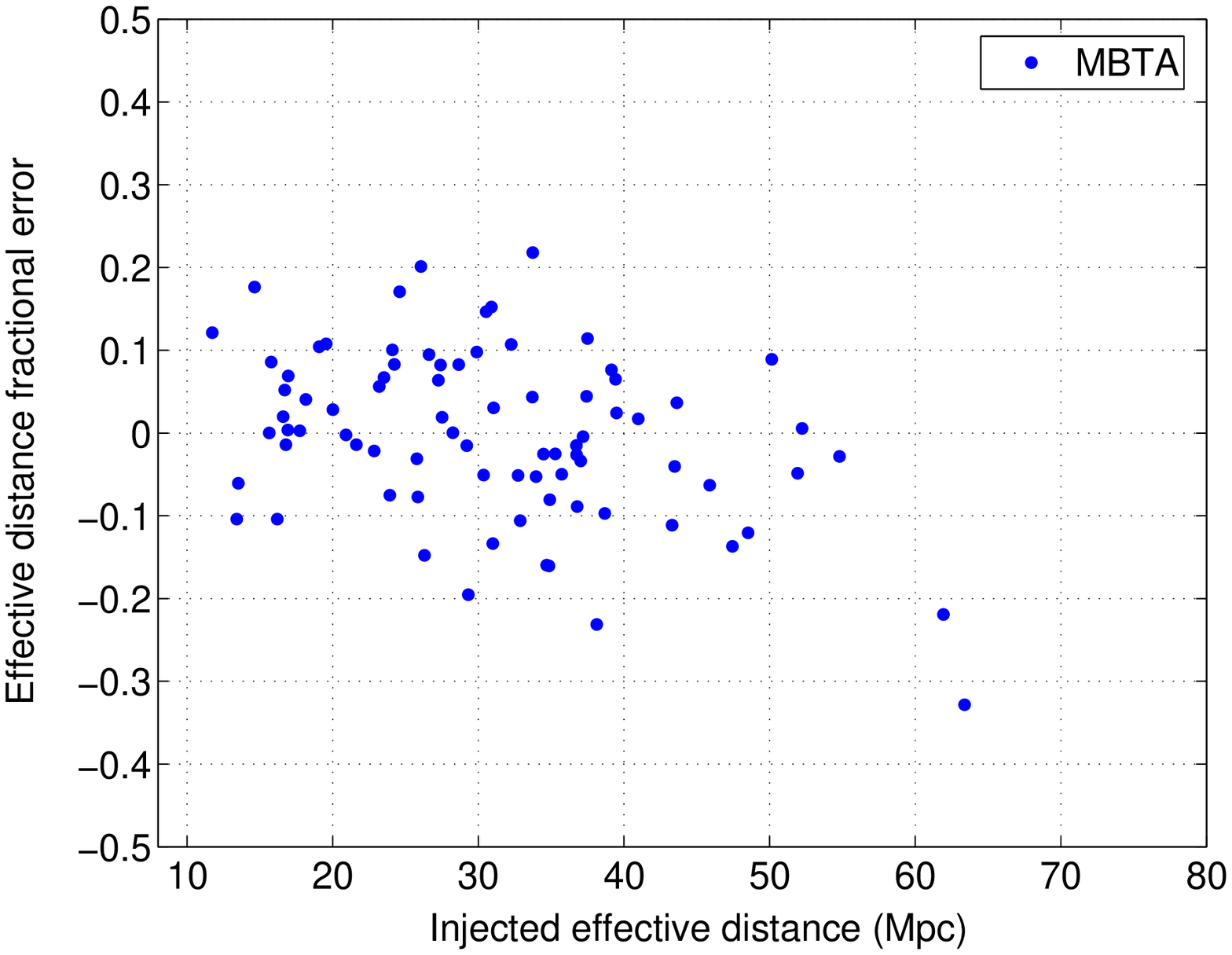}
    \includegraphics[width=2.6in,angle=0]{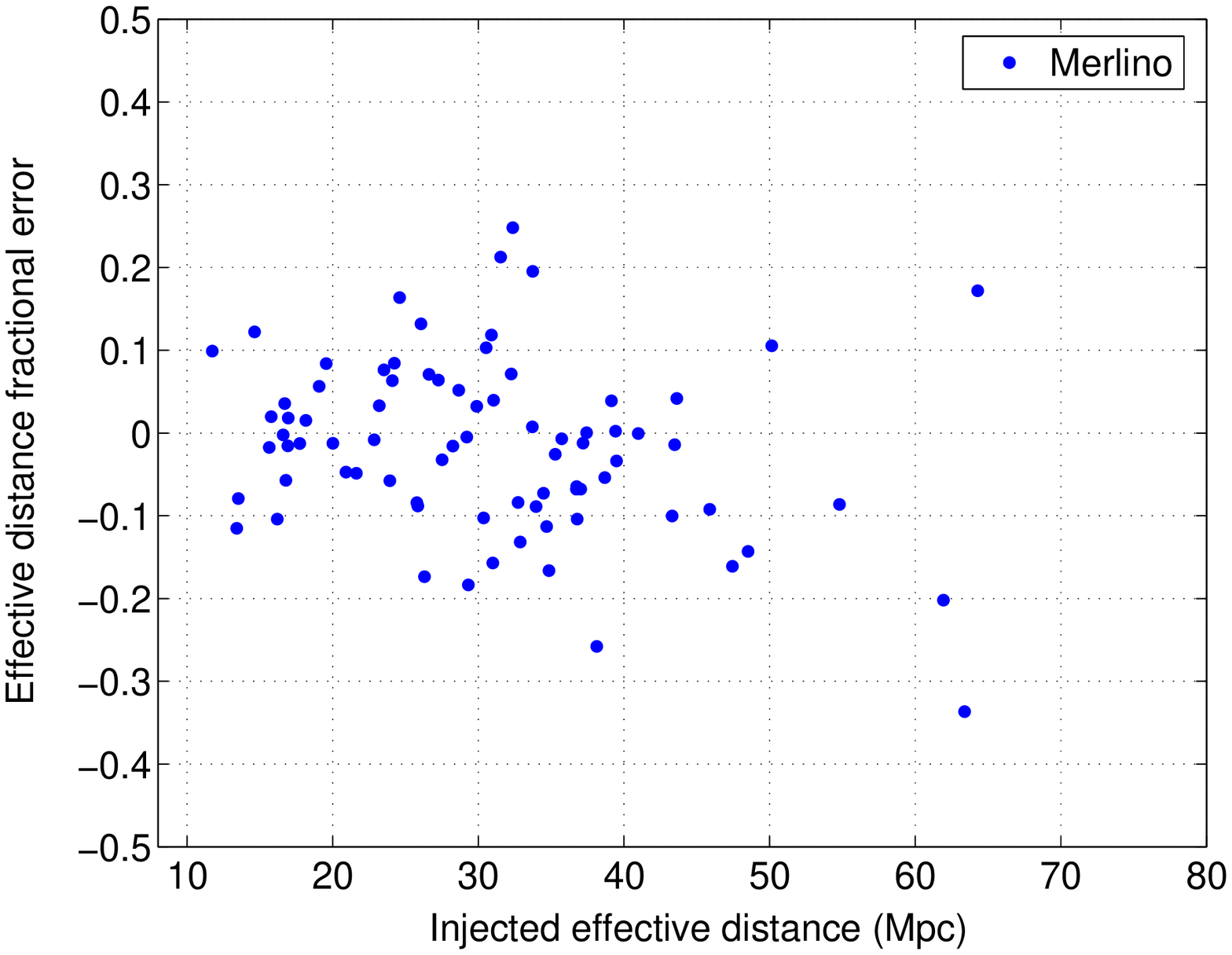}
      \includegraphics[width=2.6in,angle=0]{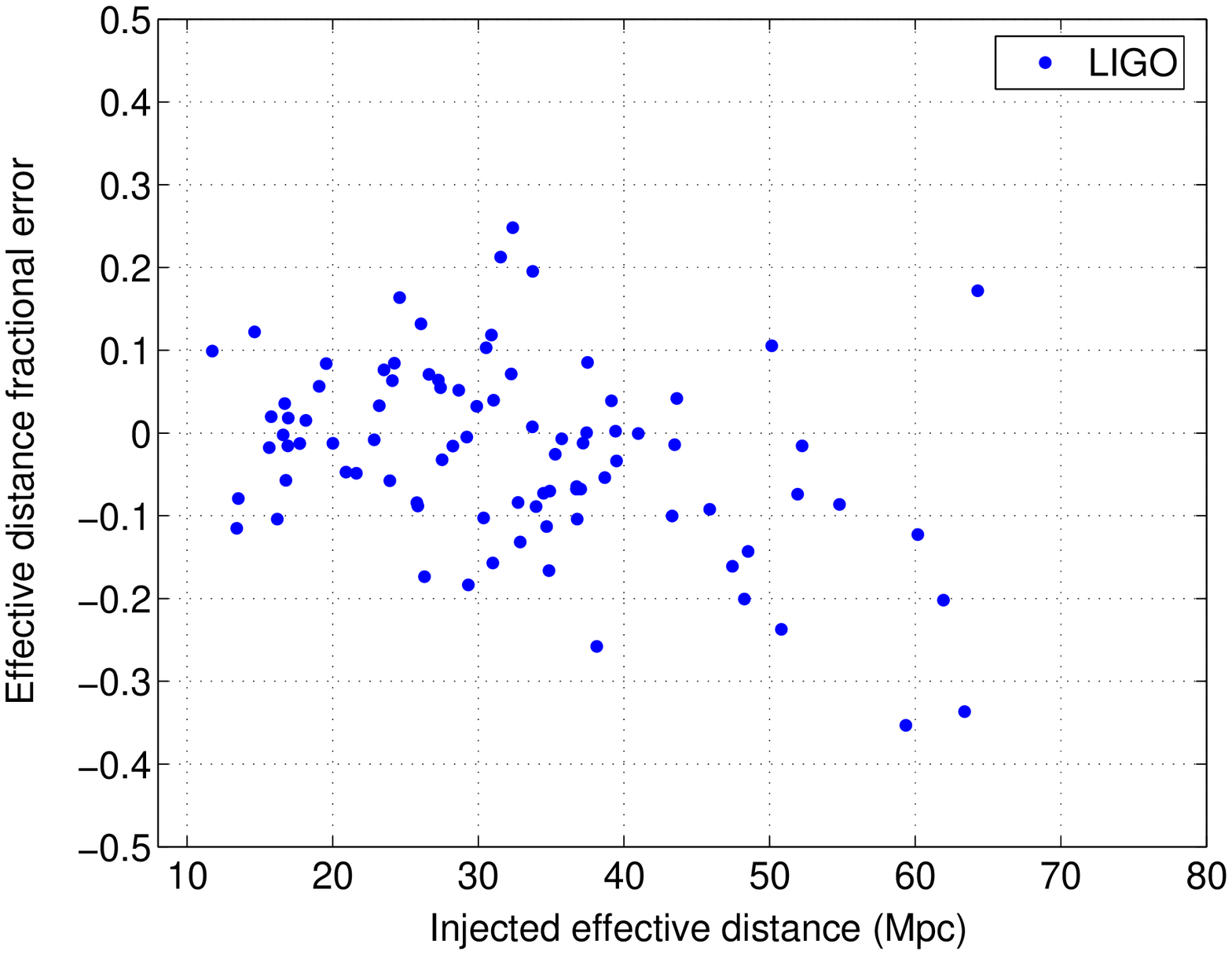}
  \end{center}
\caption{The effective distance determination accuracy as a function of actual effective distance, for the Virgo MBTA, Virgo Merlino, and LIGO inspiral analysis pipelines. These three pipelines were applied to the Virgo data, and displayed are the results from the signals recovered. The effective distance fractional error is defined as (recovered effective distance - actual effective distance)/(actual effective distance).}
\label{Vdist2}
\end{figure}

The ability to quantify the accuracy with which we can recover various injected parameters was an important goal for this study of the inspiral detection pipelines.  Using the Virgo signal data histograms were created for the parameter determination difference, with Fig.~\ref{Vchirp} for the chirp mass, and Fig.~\ref{Vend} for the end time. The difference is defined as (recovered - injected).
Fig.~\ref{Vdist} displays the effective distance parameter determination in terms of fractional error, while Fig.~\ref{Vdist2} displays the fractional error in the detected effective distance versus the actual effective distance.  The effective distance fractional error is defined as (recovered effective distance - actual effective distance)/(actual effective distance). In Fig.~\ref{Vdist2} one can see that for large injected distances, the recovered distance tends to be less than the injected distance; this is because these events are near to the threshold of the search. If the noise acts to make the signal weaker (i.e. to increase the effective distance) the event will not be detected. However, if the noise acts to make the signal stronger (i.e. to decrease the effective distance) the pipeline will detect the injected signal above threshold; this effect was observed in all the inspiral detection pipelines that we studied.

\subsection{Results for Virgo flat band search pipeline of inspiral signals}
\label{Vflat}
One of the main goals for Virgo is the realization of a reliable real time
observation strategy in order to use the interferometer as a gravitational wave observatory. With this aim, the computational strategy was carefully designed for a binary inspiral search by addressing the computational size of the problem, available computational resources, communication requirements, data handling, and the constraints due to real-time analysis conditions. The Distributed Signal
Analyzer (DiSA)~\cite{Merlinop}, named Merlino, is a particular binary inspiral search solution implemented in Virgo, based on a
parallel-distributed applications environment. This framework is composed of several processes communicating via Message Passing Interface (MPI), distributing and controlling user algorithms and the data.
The algorithms can be dynamically changed and inserted in the Merlino logic
data flow using a plug-in strategy. Specifically, the coalescing binaries plug-ins have been used to analyze the data examined in the study presented in this article; the overlap-add data handling method and the storage in memory of the template bank are implemented in order to further speed-up the analysis~\cite{Merlinop}.

A total of 6800 templates were used to cover the $1.0 M_\odot$ to $3.0 M_\odot$ range in the Virgo data, while 2000 templates were used for the LIGO data; the template bank had a minimal match of 0.95. A $\chi^2$ threshold~\cite{Allen} with 15 bands was used, and triggers were required to have a $SNR > 6$. All events within $\pm 10$ ms were clustered, and the event with the largest SNR that satisfied the $\chi^2$ cut was selected as the trigger; that is, if this trigger was within $\pm 10$ ms of an injection event end time then it was specified as a detection. This flat search code found 55\% of the binary inspiral signals injected into the V1 data, 59\% of the signals in the H1 data, and 55\% of the signals in the L1 data. The false alarm trigger rate was 0.1 Hz for the Virgo data, and 0.03 Hz for the LIGO data.

The performance of the Virgo Merlino pipeline to resolve signal parameters was comparable to the other pipelines, and example results are displayed here for the analysis of the Virgo data set. Fig.~\ref{Vchirp} displays the ability of the Merlino code to accurately determine the chirp mass. The accuracy of the end-time parameter is presented in Fig.~\ref{Vend}, while Fig.~\ref{Vdist} displays the effective distance parameter estimation.  Fig.~\ref{Vdist2} also displays the fractional error in the detected effective distance versus the actual effective distance.

\subsection{Results for LIGO inspiral detection routine}
\label{LIGO}

The LIGO inspiral detection pipeline has been used to search for signals and set upper limits with data from LIGO's first two scientific data runs~\cite{LIGO-IN,LIGO-IN2}. These publications also present detailed descriptions of the LIGO inspiral pipeline. This same LIGO inspiral pipeline was applied to the 24 hours of data for this present study. The simulated signals from H1, L1 and V1 were all examined with the LIGO code.

For the initial single detector test a threshold of $SNR > 6$ was used, with no $\chi^2$ threshold, or mass consistency check. The data was high-pass filtered (including the injections) for each interferometer individually. The template bank spanned the range from $1 M_\odot$ to $3 M_\odot$ and had a minimal match of 0.95; a total of 10900 templates were used to analyze the Virgo data set, and 2900 templates for the LIGO data. When calculating the detection efficiency we required that a candidate trigger occur within $\pm 10$ ms of the injected signal's end-time. There was no clustering over the template bank. For the H1 data 66\% of the inspiral signals were detected, while the pipeline found 64\% of the L1 signals. The efficiency of signal detection efficiency was 62\% for the V1 data. For the LIGO pipeline the false alarm trigger event rate was 0.07 Hz when analyzing the LIGO data and 0.77 Hz for the Virgo data set.

 The recovered parameter values correspond to the template producing the largest SNR trigger within 10 ms of the actual end time. Using the Virgo signal data, we created histograms of the parameter determination difference (Fig.~\ref{Vchirp} for the chirp mass, Fig.~\ref{Vend} for the end time, and Fig.~\ref{Vdist} for the effective distance), as well as a plot, Fig.~\ref{Vdist2} that displays the fractional error in the detected effective distance versus the actual effective distance.

\section{MCMC Parameter Estimation}
\label{MCMC}
Parameter estimation, and the generation of a posterior probability density function (PDF) for each parameter, was also done utilizing a Markov chain Monte Carlo (MCMC) routine. These MCMC methods are part the LIGO binary inspiral data analysis effort. The basic operation of the inspiral MCMC code is described in~\cite{NCMCMC}, which also contains a description of MCMC techniques. The purpose of this code is to take triggers generated from the LIGO inspiral pipeline, and then examine that section of the data about the event. This MCMC code was applied to those sections of data in this study where coincident events were found. The MCMC is not a search pipeline, as the program is too computationally taxing to be applied to all of the data. The MCMC searched for events that had a binary coalescence end-time within a $\pm 50$ ms window of coincident triggers from the LIGO inspiral search pipeline.

The MCMC code used in the present study was written in C, and looked for inspiral signals based on 2.0 PN signals in the frequency domain. 
For the present study the {\it prior} for the masses of the compact objects was uniform from 0.9 $M_{\odot}$ to 3.1 $M_{\odot}$ range. For this problem there are five parameters to estimate: the binary masses, $m_1$ and $m_2$, the effective distance $d_L$, the phase at coalescence $\phi_c$ and the time at coalescence $t_c$. The program reparameterizes the masses in terms of the chirp mass $m_c=(m_1 m_2)^{3/5} /(m_1+m_2)^{1/5}$ and the mass ratio parameter $\eta=m_1 m_2/(m_1+m_2)^2$.
This inspiral MCMC code begins with a method called {\it importance resampling}~\cite{Gel97, Gel98}; at the start the code first generates a large sample of parameter space points from a
distribution covering the whole prior, and then draws the actual sample out of these
with correspondingly assigned weights depending on the posterior density. The Markov chains are started in regions of parameter space that are {\it likely} to be close to the true parameter values. 
 Simulated annealing~\cite{Gilks96} was used to optimize the initial {\it burn-in} of the Markov chains. During the burn-in period, the effect of the noise in the likelihood function is arbitrarily increased (an effective temperature increase); this simulated annealing technique was introduced in~\cite{Metropolis53} and allows scanning of the whole parameter space by permitting larger steps.
The candidate generating function~\cite{NCMCMC} for the parameters is designed so that correlations between chains are measured as the program progresses, and correlation values are fed back to the generating function. 

\begin{figure}[tb]
  \begin{center}
    \includegraphics[width=3.0in,angle=-90]{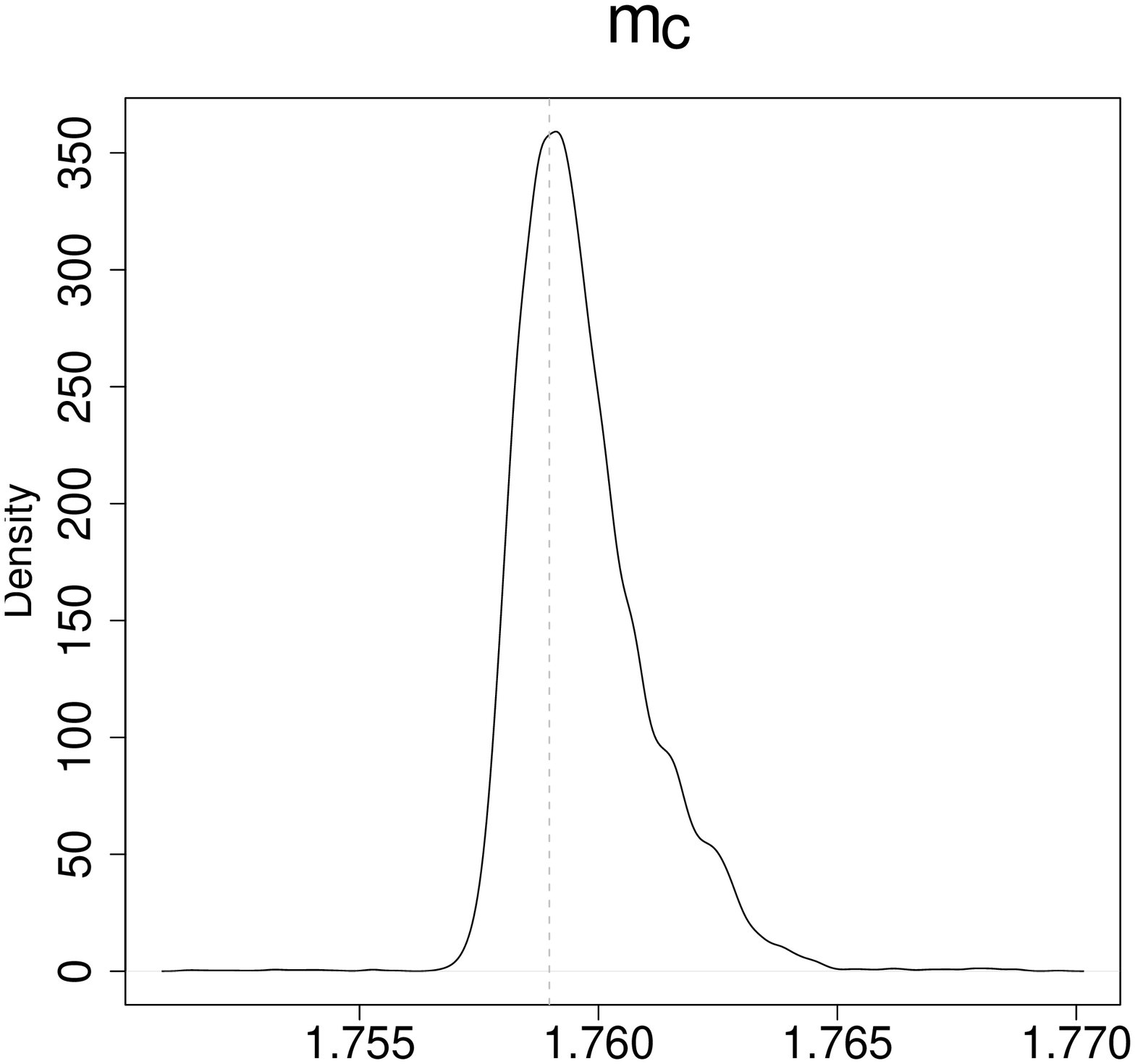}
    \includegraphics[width=3.0in,angle=-90]{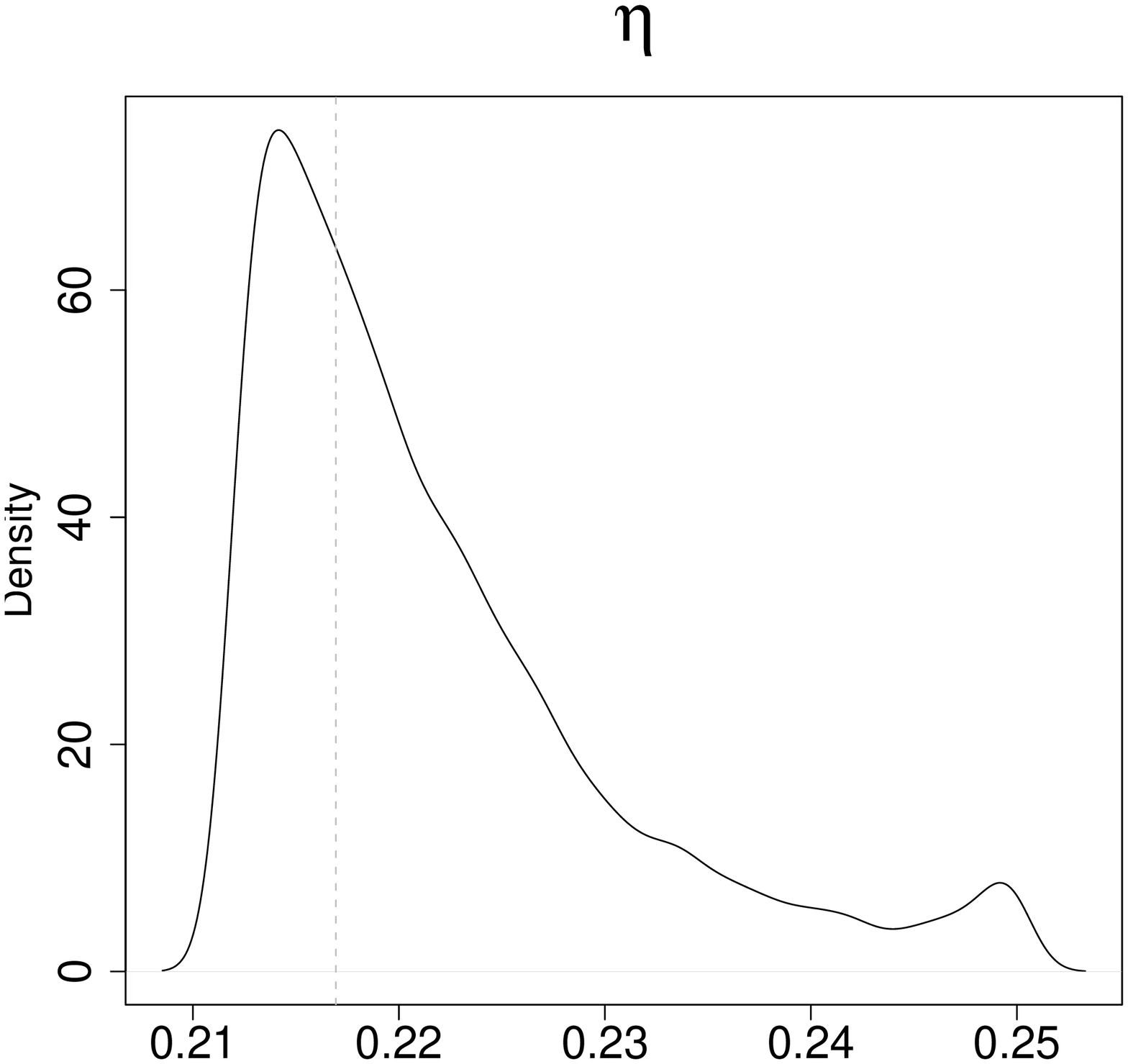}
    \includegraphics[width=3.0in,angle=-90]{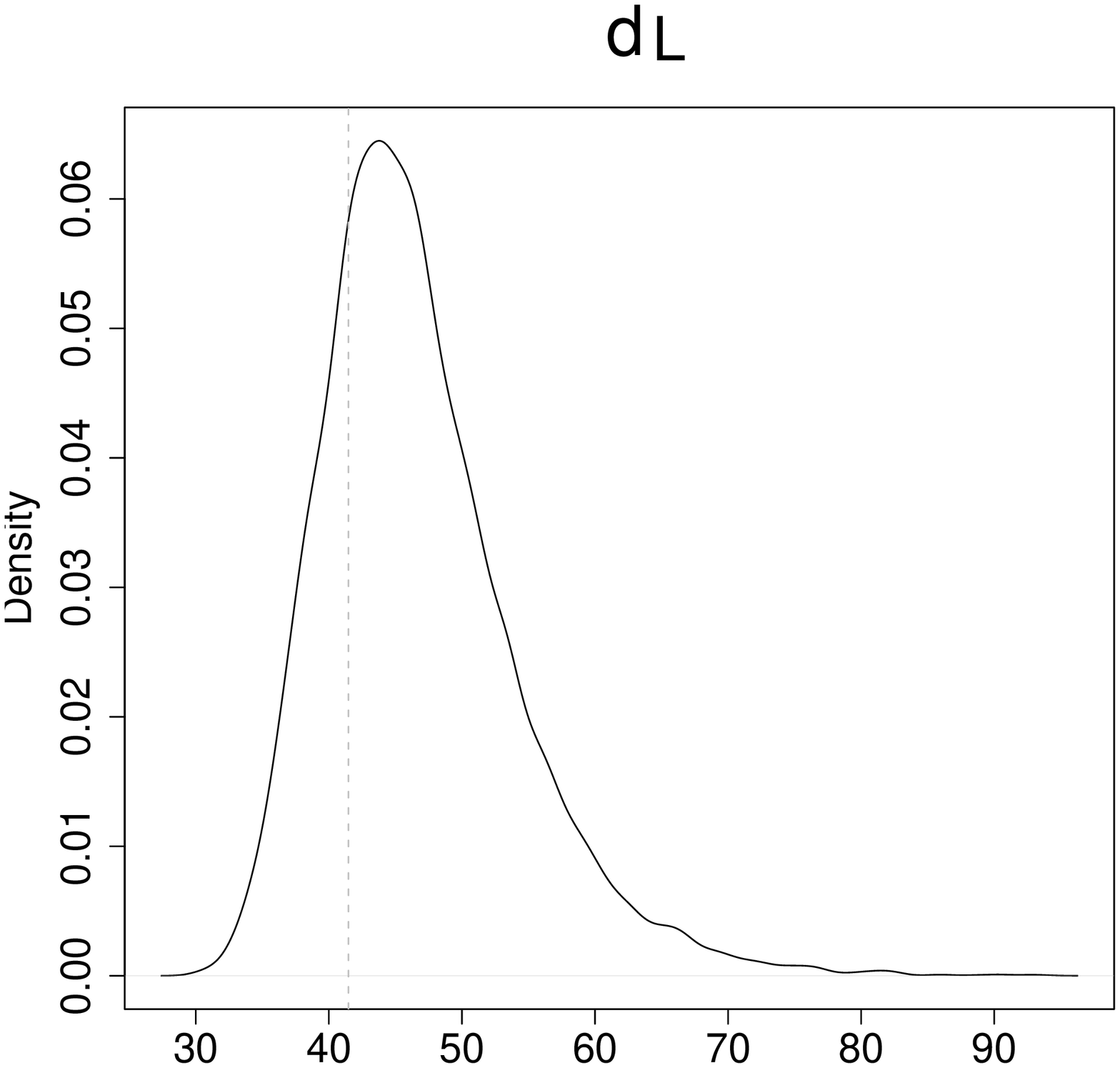}
    \includegraphics[width=3.0in,angle=-90]{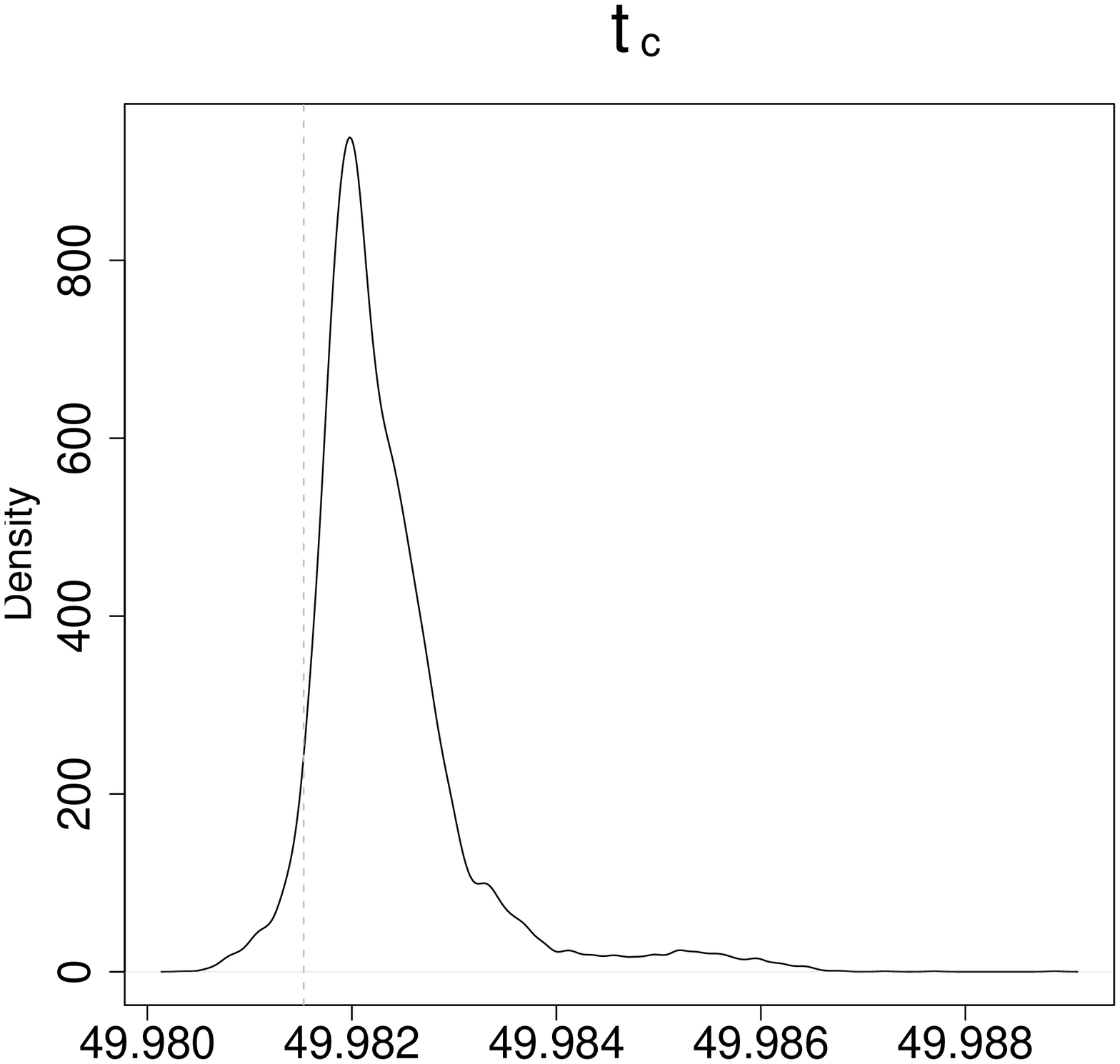}
  \end{center}
\caption{MCMC produced posterior PDFs for chirp mass $m_c$, mass ratio $\eta$, effective distance $d_L$ and the coalescence time $t_c$ for one event. In this example the binary inspiral signal is embedded in the H1 data. The actual parameters used to produce this signal, displayed by the dashed vertical lines, are $m_1=1.4 M_{\odot}$, $m_2=3.0 M_{\odot}$ ($m_c=1.759 M_{\odot}$ and $\eta=0.217$), $d_L=41.47 Mpc$ and $t_c=49.9815 s$.}
\label{figMCMC}
\end{figure}

The MCMC parameter estimation code was applied to all of the LIGO-Virgo data from this study, and here we show examples from the analysis of the LIGO signals. Fig.~\ref{figMCMC} shows an example of MCMC generated estimates for the posterior probability distribution functions for a simulated event in the H1 data detected with the LIGO pipeline. This signal had real parameter values of  $m_1=1.4 M_{\odot}$, $m_2=3.0 M_{\odot}$ ($m_c=1.759 M_{\odot}$ and $\eta=0.217$), $d_L=41.47 Mpc$ and $t_c=49.9815 s$. From the generated posterior PDFs estimates for the parameter values can be associated with the mean of the distribution. The error in the parameter estimate would correspond to a particular width. We will give the 5 to 95 percentile range of the posterior distribution, which gives a 90\% posterior credibility interval. For this example the parameters' mean values and 90\% credibility ranges are $m_c = 1.7597~ (1.7580 - 1.7623)~ M_{\odot}$, $\eta = 0.2215~ (0.2126 - 0.2407)$, $d_L = 46.669~ (36.936 - 60.023)~ Mpc$, and $t_c = 49.9823~ (49.9815 - 49.9837)~ s$.

\begin{figure}[tb]
  \begin{center}
    \includegraphics[width=3.0in,angle=0]{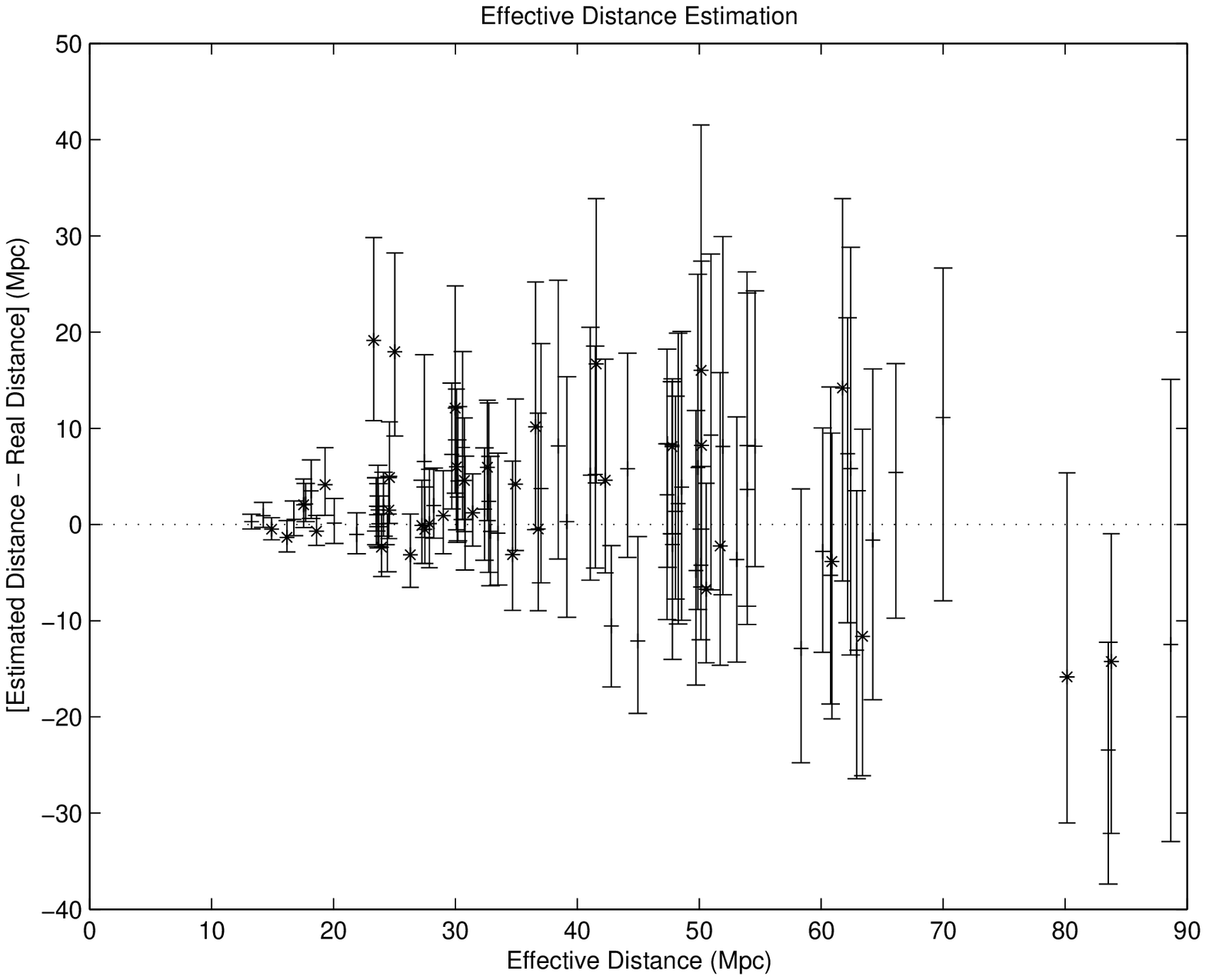}
    \includegraphics[width=3.0in,angle=0]{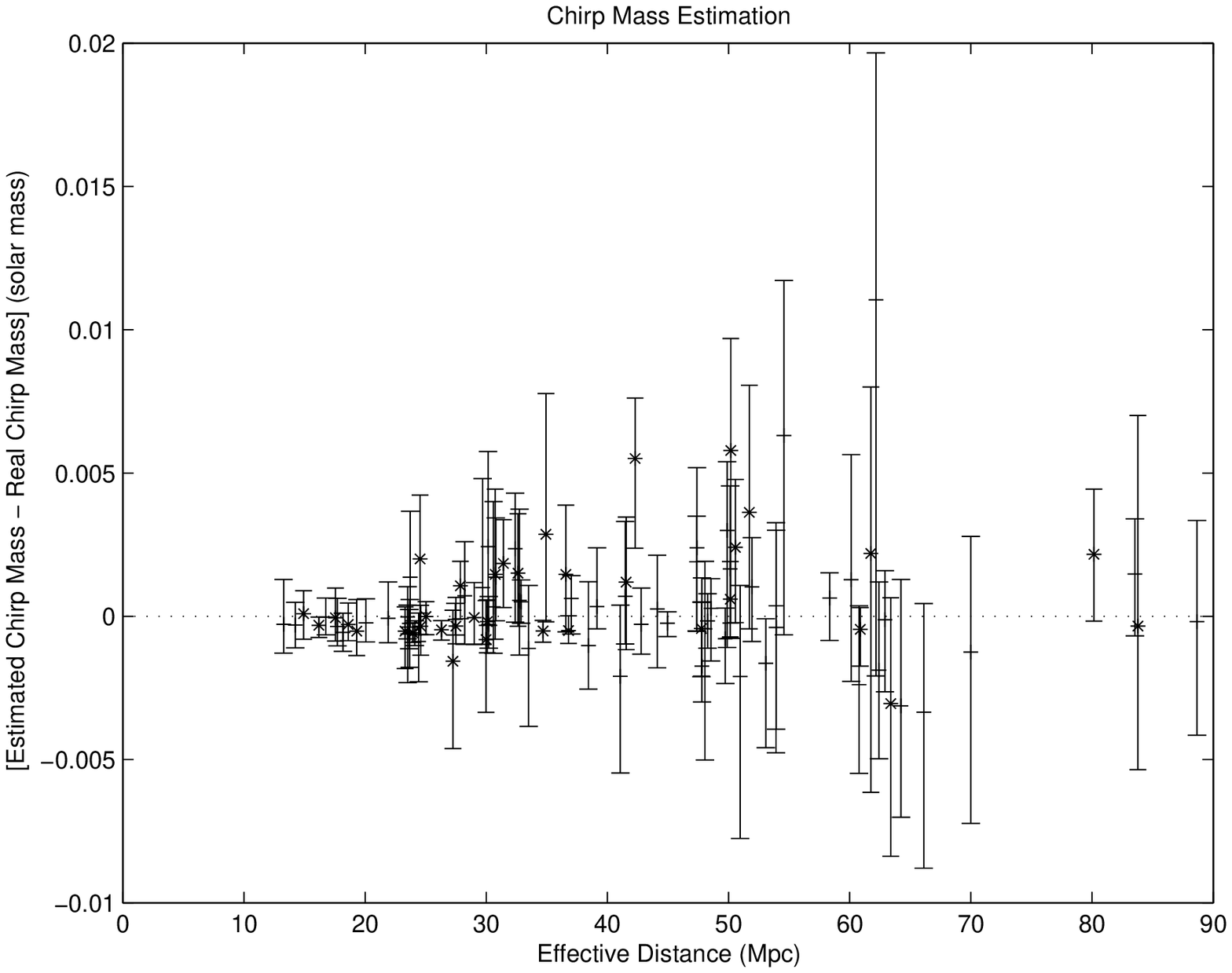}
    \includegraphics[width=3.0in,angle=0]{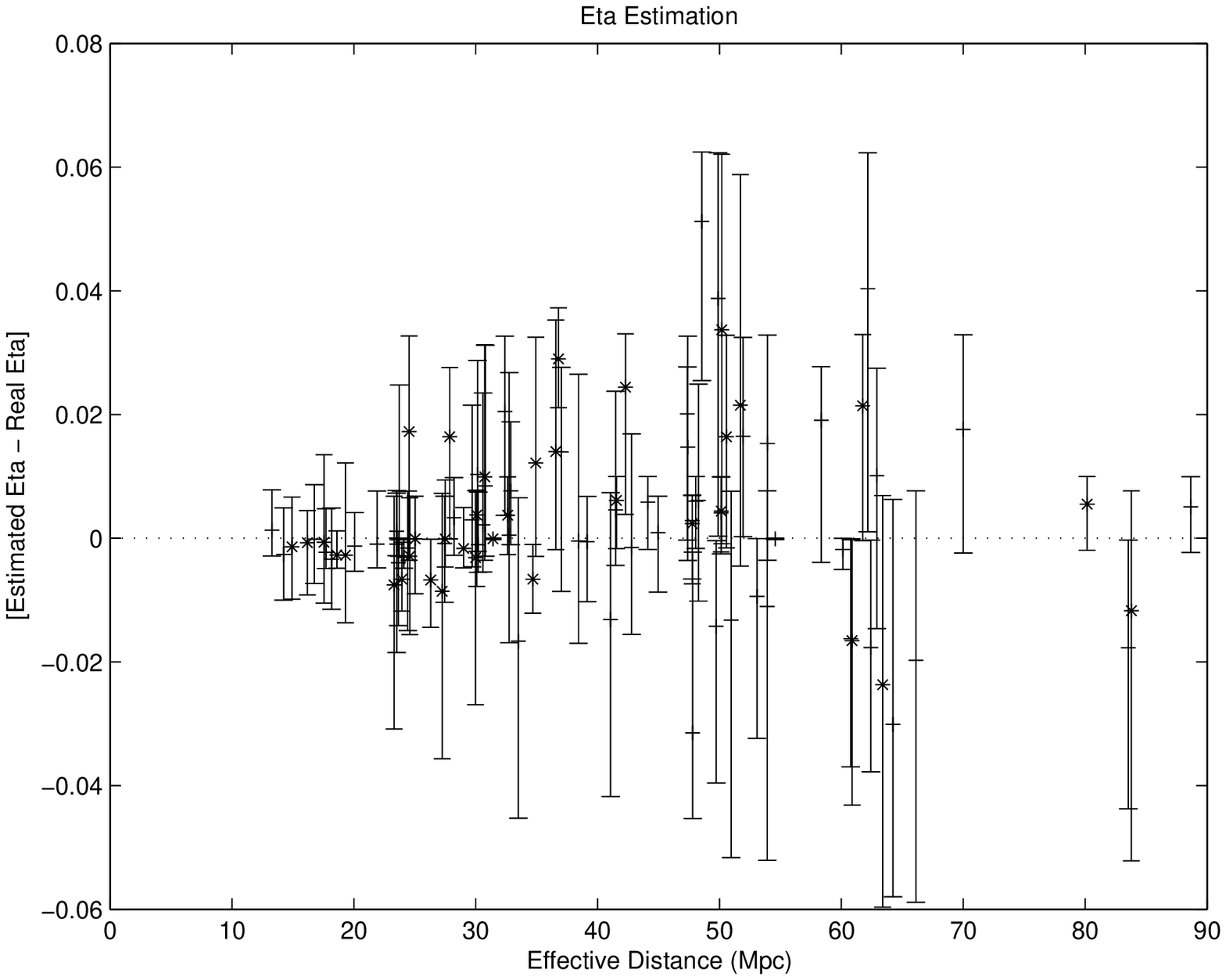}
    \includegraphics[width=3.0in,angle=0]{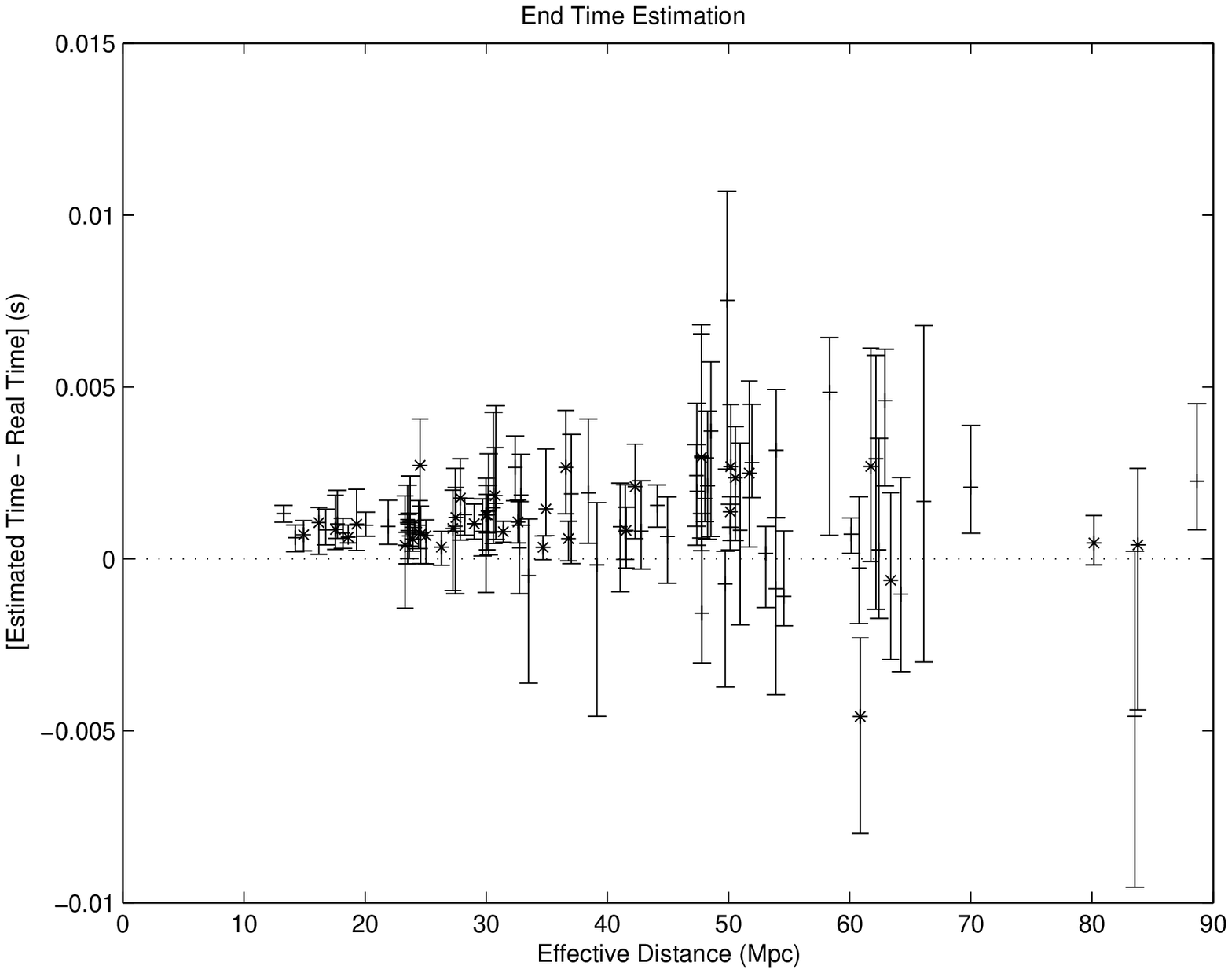}
  \end{center}
\caption{The parameter estimation accuracy for the MCMC chirp mass $m_c$, mass ratio $\eta$, effective distance $d_L$ and the coalescence time $t_c$ are shown. Displayed are the differences between the MCMC parameter estimated and the actual value, as a function of effective distance. These events were from the H1 and L1 data. The error bars correspond to the 90\% credibility ranges.}
\label{figMCMC2}
\end{figure}

As an example of the parameter estimation accuracy, Fig.~\ref{figMCMC2} shows the difference between the MCMC estimated parameter values and the real values, as a function of effective distance. The errors represent the 90\% credibility ranges. These events were from the simulated inspiral signals found by the LIGO pipeline in both the L1 and H1 data sets. All mass combinations are represented in this plot.

\section{LIGO and Virgo Inspiral Detection Pipeline Comparison}
\label{comparison}
The Merlino, MBTA and LIGO binary inspiral pipelines all operated in a comparable way, and produced good single interferometer detection statistics. Triggers were recorded, and those with the largest $SNR$ had the parameter values for the templates noted. The ability to resolve the parameters was equally positive (within statistical errors and other uncertainties) for all of the pipelines. A major conclusion from our study is that the three inspiral search pipelines performed equally well, and there is mutual confidence between the groups in the others' inspiral search abilities. The comparisons presented in this section were all conducted using the Virgo data.

\begin{figure}[tb]
  \begin{center}
    \includegraphics[width=2.6in,angle=0]{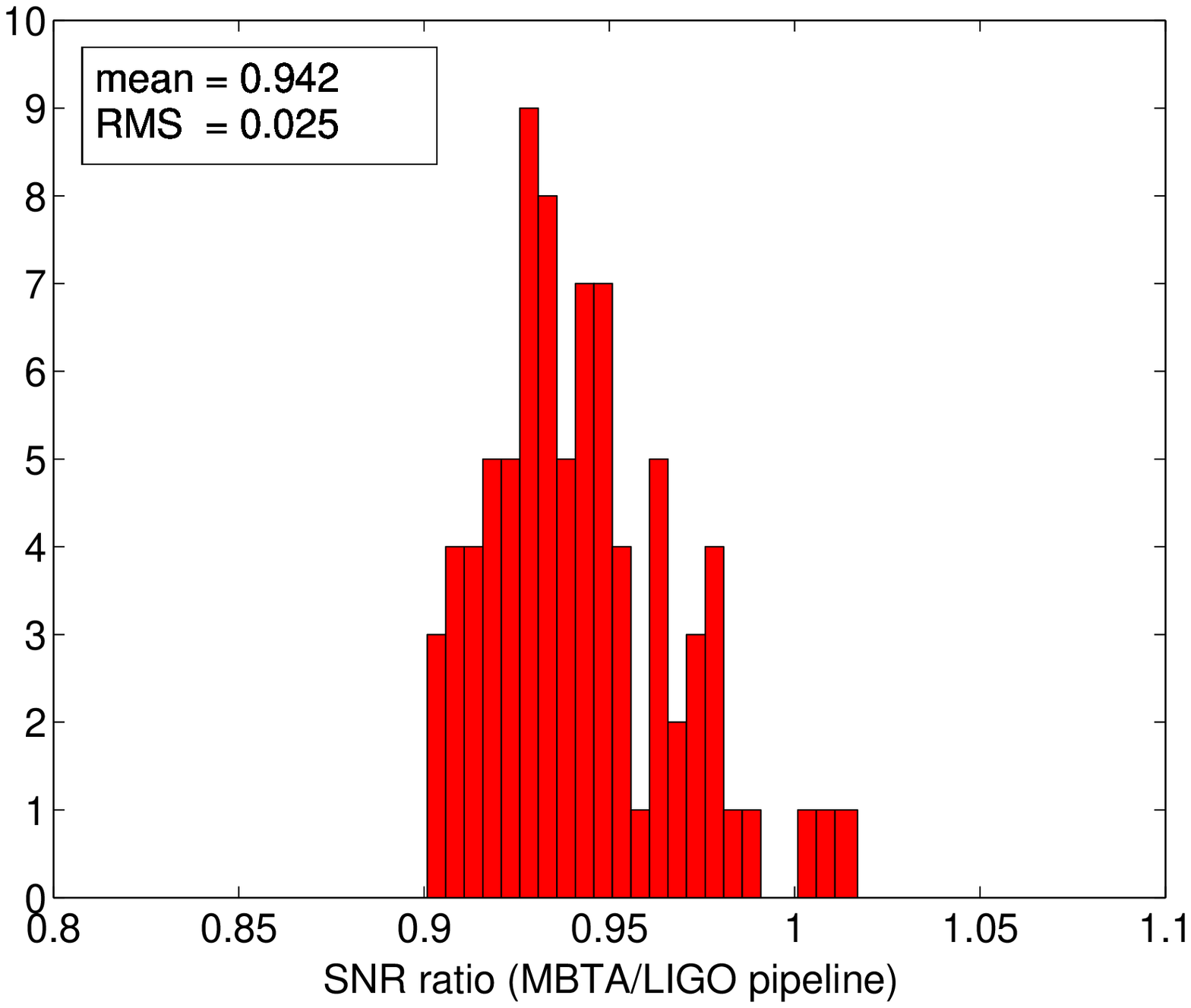}
    \includegraphics[width=2.6in,angle=0]{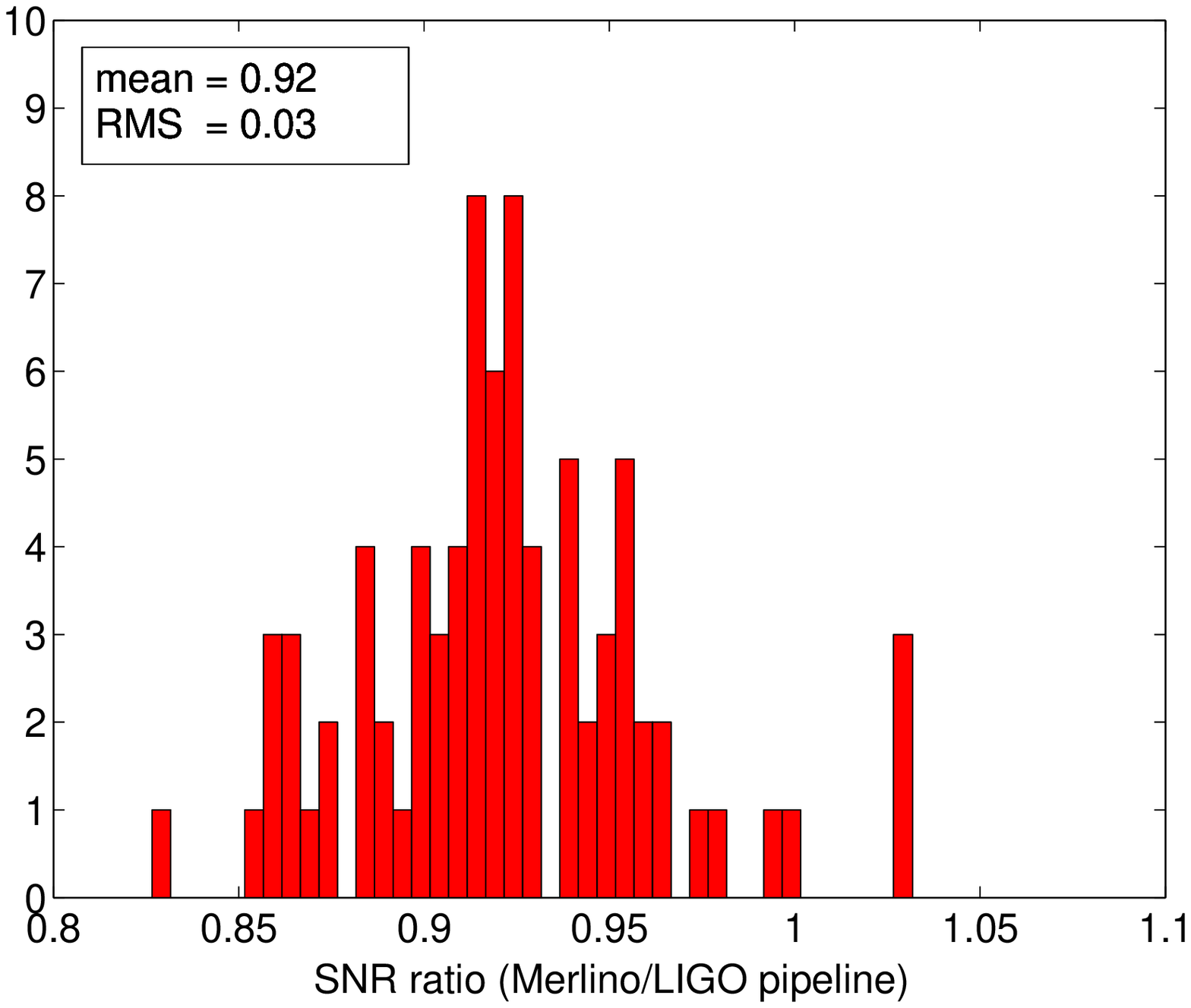}
  \end{center}
\caption{On the left is a histogram of the ratios of the SNRs for MBTA and LIGO events detected by both pipelines in the V1 data. The mean and RMS values for the distribution are given in the figure, showing a 6\% excess in the size of the LIGO SNR. On the right is a histogram of the ratios of the SNRs for Merlino and LIGO events detected by both pipelines in the V1 data; there is a 8\% excess in the size of the LIGO SNR.}
\label{compMBTA}
\end{figure}

In order to display the output of the pipelines, we present here a direct comparison of the Merlino, MBTA and LIGO results from the V1 data. Fig.~\ref{compMBTA} (left) shows a histogram of the ratio of the MBTA SNR to the LIGO SNR for the events detected. Similarly, Fig.~\ref{compMBTA} (right) shows a histogram of the ratio of the Merlino SNR to the LIGO SNR for the events detected. There were some slight differences in the results between the pipelines. For example, the SNR of detected events was about 6\% larger from the LIGO pipeline versus those from the MBTA pipeline, and about 8\% larger than those from the Merlino pipeline.

The difference in the SNRs is predominantly affected by slightly differing
methods for calculating the noise power spectral density (PSD).
Specifically, the LIGO pipeline calculates its PSD via the median power
spectrum (the median is calculated frequency-bin by frequency-bin) of 15
overlapping segments, where each segment was 256 s in length. The median
was chosen for the LIGO pipeline PSD generation so that it would not be
overly biased by a single loud glitch (as could happen when using the
mean), at the cost of a larger bias in the case of gaussian noise. The PSD
calculated by the Virgo pipeline is computed from a larger number of
averages, as the mean from 1800 s of data, each section 16.38 s long, with
no overlap.  We conducted a direct comparison of the PSDs generated by the
LIGO and Virgo pipelines. The estimate of the $SNR^2$ scales with
frequency like $f^{-7/3}/S(f)$, where $S(f)$ is the noise PSD. When we sum
$f^{-7/3}/S(f)$ over the frequencies from 40 to 2048 Hz the LIGO result
exceeds that of Virgo by 10\%, giving an overestimation in the LIGO SNR of
5\%. Numerical experiments have subsequently verified that increasing the
number of segments used in calculating the noise PSD will reduce the bias.
It should be noted that the bias from the noise PSD strictly cancels out
in the estimation of the effective distance.


We were also concerned that the frequency domain stationary phase approximation used by LIGO could affect the SNR estimation. Specifically, the signals for this study were generated to 2.0 PN order in the time domain. For both Virgo pipelines the frequency domain detection templates are the Fourier transforms of the time domain 2.0 PN signals. On the other hand, the LIGO pipeline uses 2.0 PN frequency domain templates generated via a stationary phase approximation. We studied the overlap between Fourier transforms of 2.0 PN time domain signals and the stationary phase frequency domain signals. Our results indicate that the SNR for the LIGO would be reduced by at most 1\% (and only for signals from the highest mass pairs). In addition we verified the {\it effectualness} and {\it faithfullness} of the signals used in this study, similar to what was done in~\cite{Damour}. Specifically for the mass pairs used in this study, the faithfulness varies between 96\% and 98\%. Lower values correspond to the larger mass ratios. The effectualness is always above 99\%, except for the $3 M_\odot$ - $3 M_\odot$ mass pair, for which the effectualness is 98.8\%. The accuracy in the estimation of the chirp mass using the stationary phase templates ranges from a few parts in $10^{-5}$ to a few parts in $10^{-4}$.

We also examined how template placement in the $m_1$ versus $m_2$ plane grid would affect the recovered SNR, and we found that for our studies this had a small effect on the SNR difference. The density of the template grid is higher for the LIGO pipeline than the Virgo pipelines. Specifics of the template placement for the inspiral pipelines used in our study are presented in~\cite{LV1I}. When examining the SNR ratio from the MBTA and LIGO pipelines we found the excess value from the LIGO pipeline to be consistently present for the cases when the two masses were equal, or when they had a relatively large difference. The increased density of the LIGO grid would give an excess in SNR, but for the study presented here it appears that the effect would not produce a SNR difference exceeding 1\%. 

From these studies we believe that we understand how differences are created in the SNR of the inspiral triggers by the LIGO and Virgo pipelines. The slightly higher detection efficiency produced by the LIGO pipeline is a direct consequence of its elevated value of the produced SNR ratio, which in turn is due to the difference in the methods used to estimate the noise PSD. All three pipelines in this study used a $SNR>6$ threshold. Those events near this cutoff would have a slight preference of being seen by the LIGO pipeline. When one accounts for the SNR artifact the detection efficiencies of all of the pipelines are seen to be the same.

\begin{figure}[tb]
  \begin{center}
    \includegraphics[width=2.6in,angle=0]{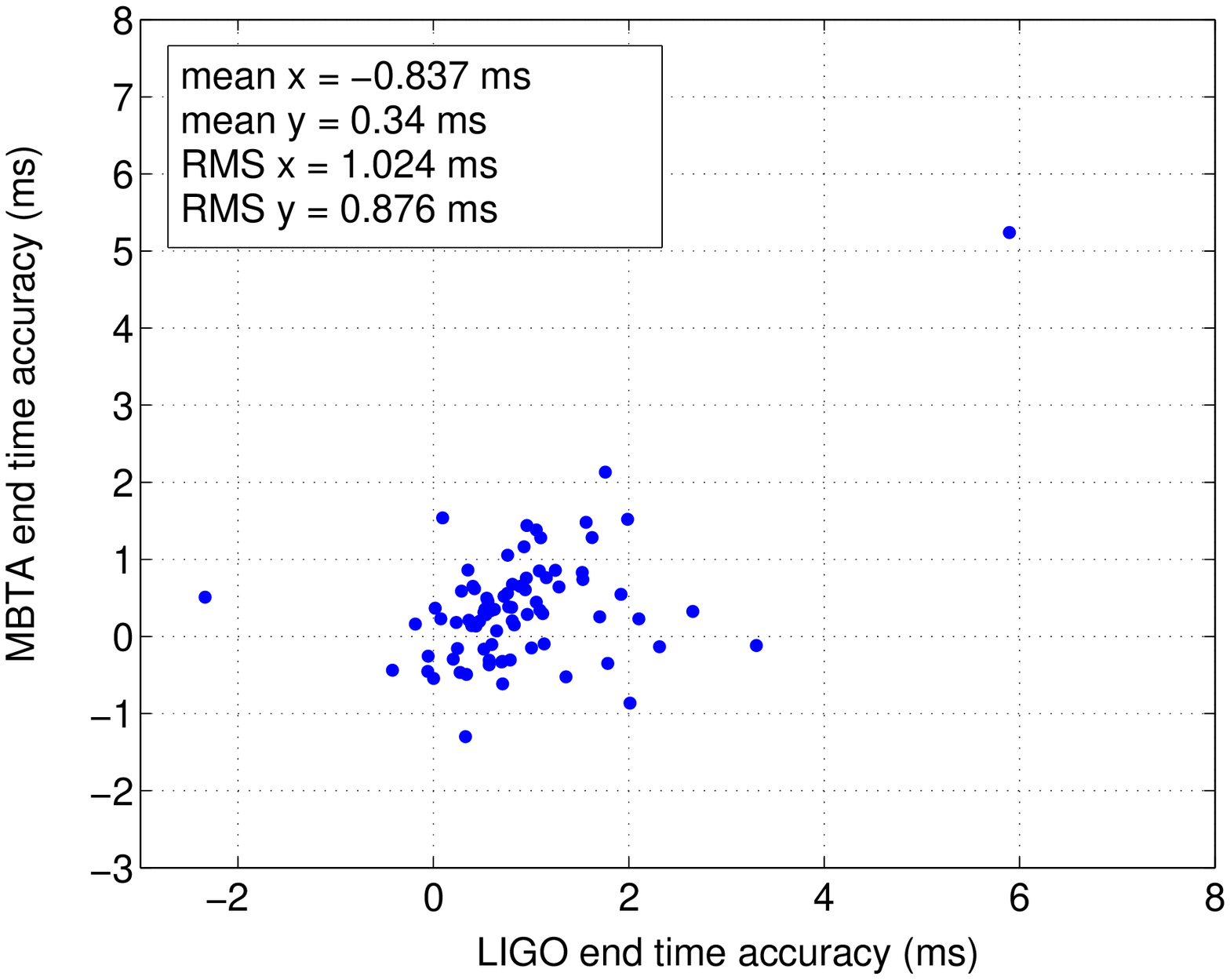}
    \includegraphics[width=2.6in,angle=0]{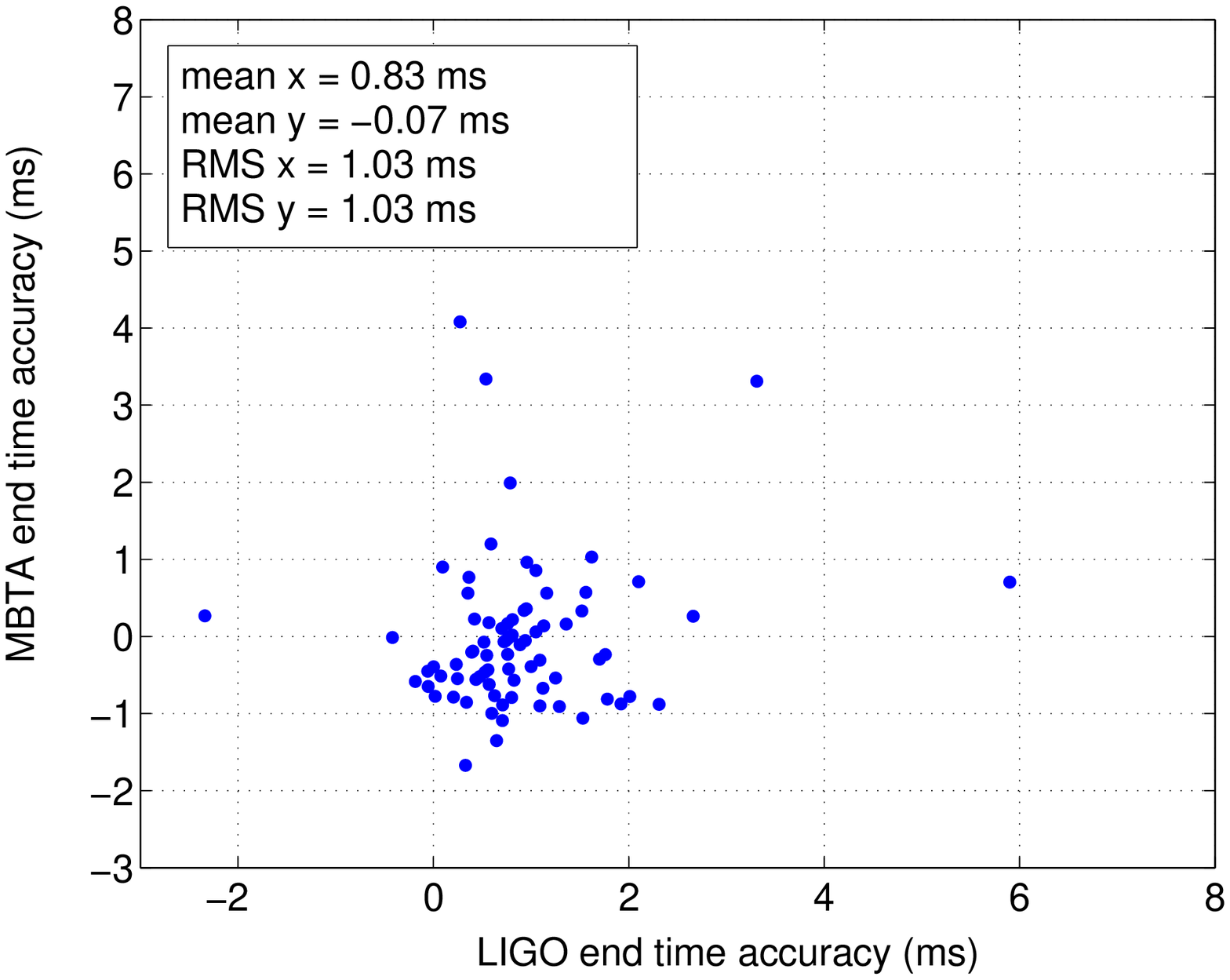}
  \end{center}
\caption{On the left is a scatter plot of the end time accuracy for MBTA and LIGO events detected by both pipelines in the V1 data. The accuracy is defined as the actual end time subtracted from the recovered end time. The mean and RMS values are given in the figure, and show that the LIGO inspiral pipeline tends to over estimate the end time by about 1 ms. On the right is a scatter plot of the end time accuracy for Merlino and LIGO events detected by both pipelines in the V1 data; again, the LIGO inspiral pipeline tends to over estimate the end time by about 1 ms.}
\label{compMBTAtime}
\end{figure}

Another small difference in the LIGO and Virgo pipelines concerns the estimate of the endtime. The LIGO pipeline typically estimates an endtime that is 1 ms off from the actual value. This can be seen in the output of the LIGO pipeline in Fig.~\ref{Vend} and the MCMC generated parameter estimates in Fig.~\ref{figMCMC2}. The 1 ms offset in the LIGO end time estimation is also apparent in Fig.~\ref{compMBTAtime} (left), where the end time accuracy for MBTA and LIGO events detected by both pipelines in the V1 data is presented. Similarly, a comparison of the LIGO and Merlino estimates of the end time also display an offset of about 1 ms for the LIGO results; see Fig.~\ref{compMBTAtime} (right).  This small difference in the nature of the detection templates (frequency domain stationary phase for LIGO, and Fourier transforms of the time domain templates for Virgo) is responsible for the small time shift.

\begin{figure}[tb]
  \begin{center}
    \includegraphics[width=2.6in,angle=0]{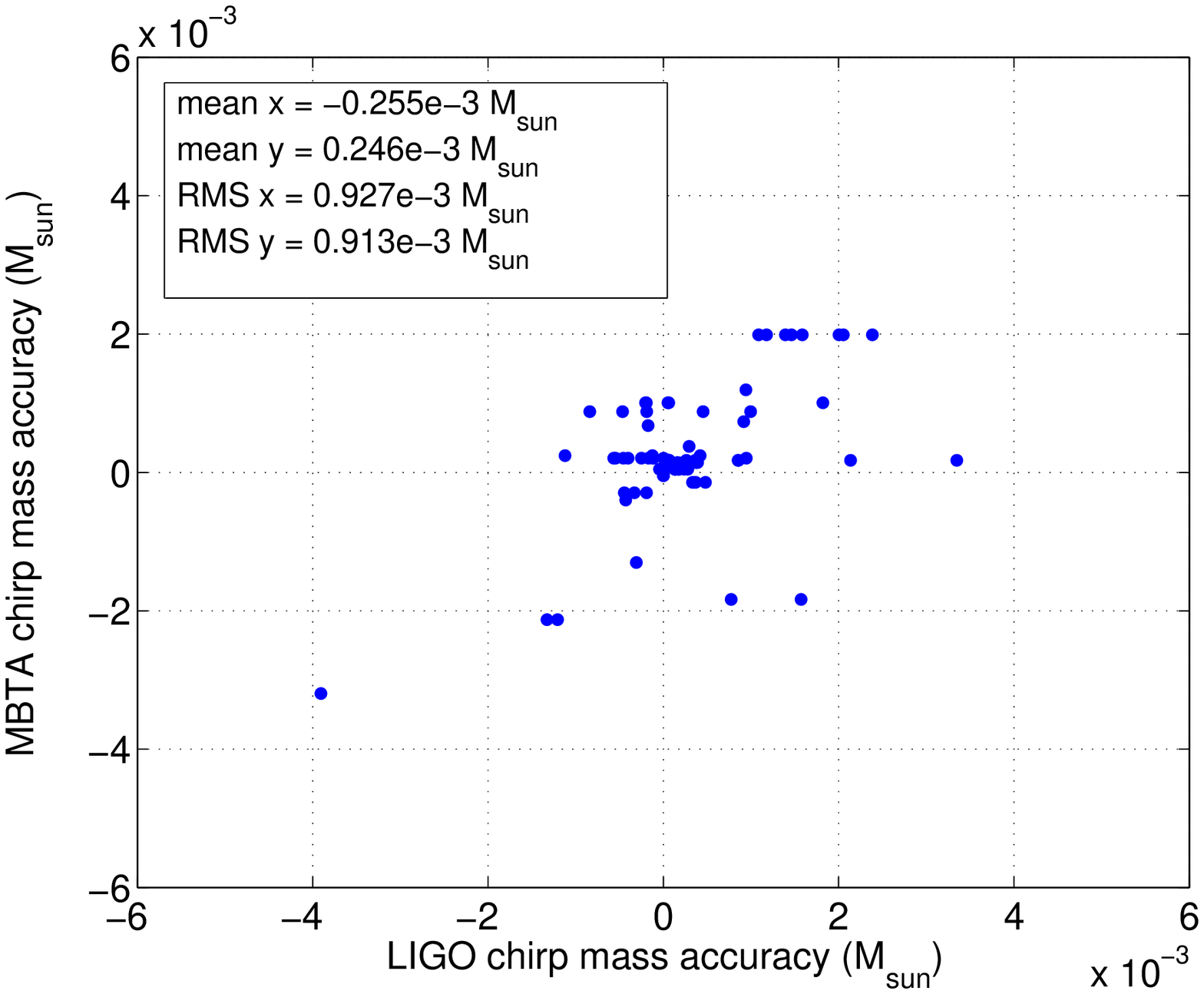}
    \includegraphics[width=2.6in,angle=0]{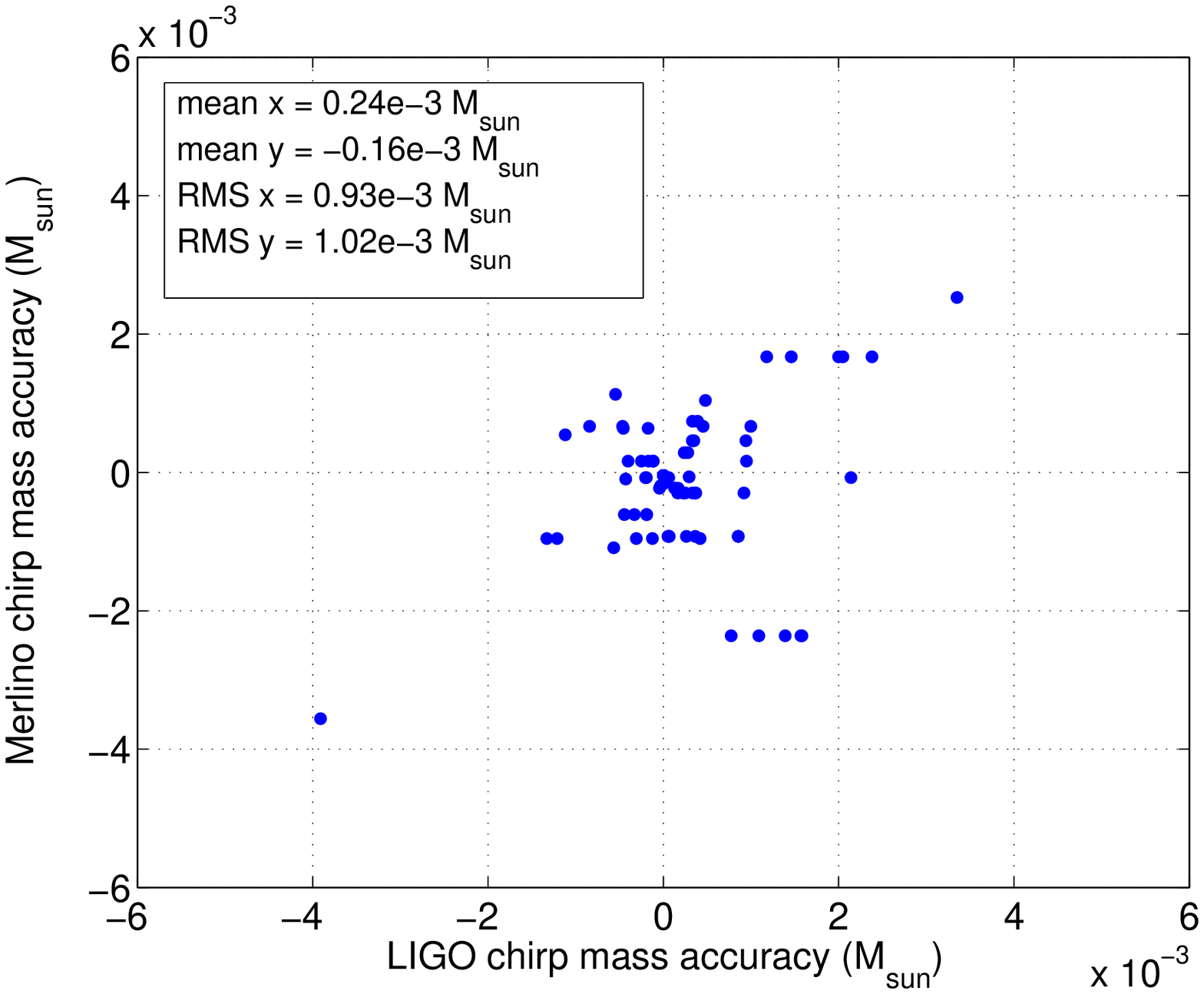}
  \end{center}
\caption{On the left is a scatter plot of the chirp mass accuracy for MBTA and LIGO events detected by both pipelines in the V1 data. On the right is a scatter plot of the chirp mass accuracy for Merlino and LIGO events detected by both pipelines in the V1 data. The accuracy is defined as the actual chirp mass subtracted from the recovered chirp mass. The mean and RMS values are given in the figures, and show that the Virgo MBTA, Virgo Merlino, and LIGO inspiral pipelines all estimated the chirp mass accurately.}
\label{compMBTAchirp}
\end{figure}

\begin{figure}[tb]
  \begin{center}
    \includegraphics[width=2.6in,angle=0]{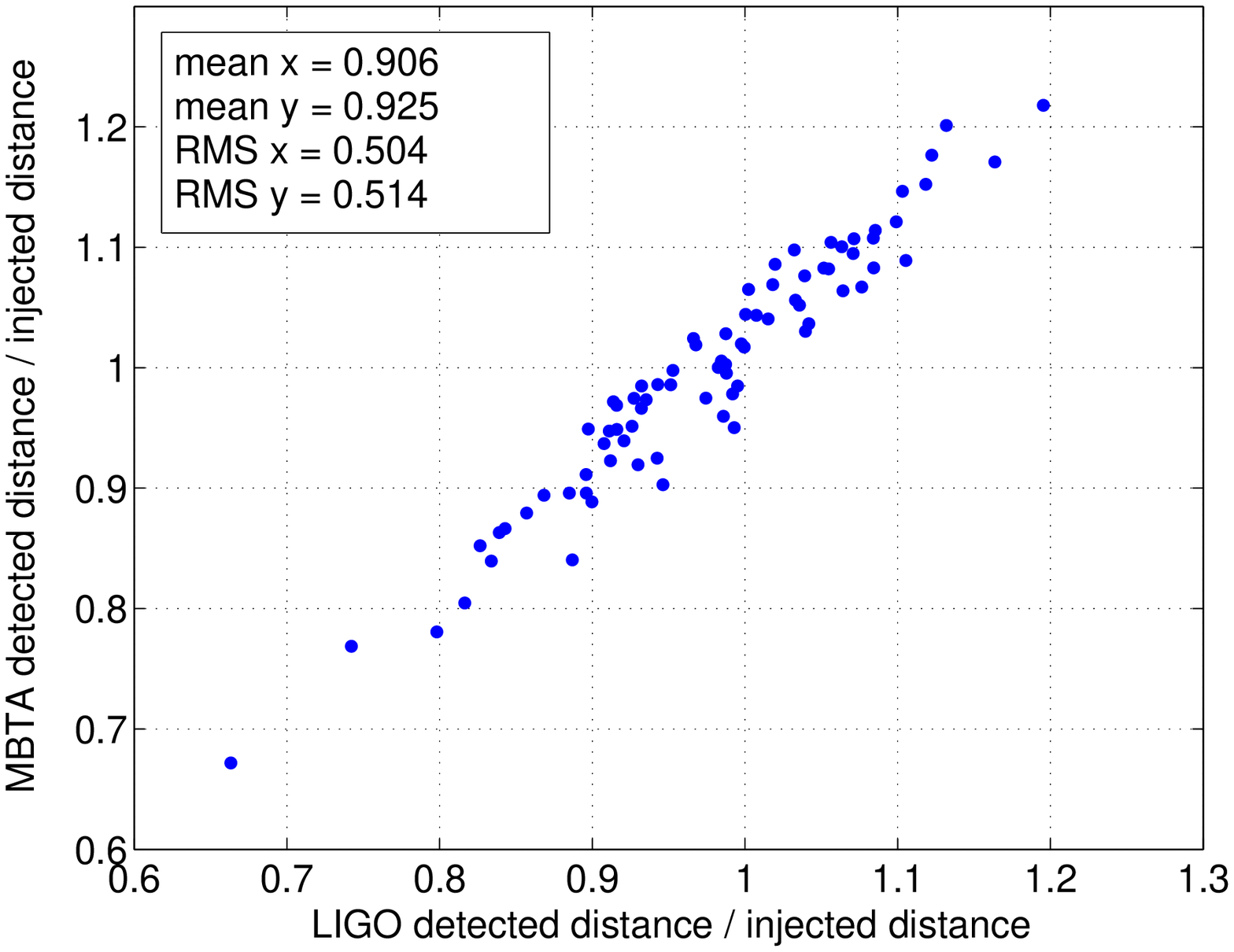}
    \includegraphics[width=2.6in,angle=0]{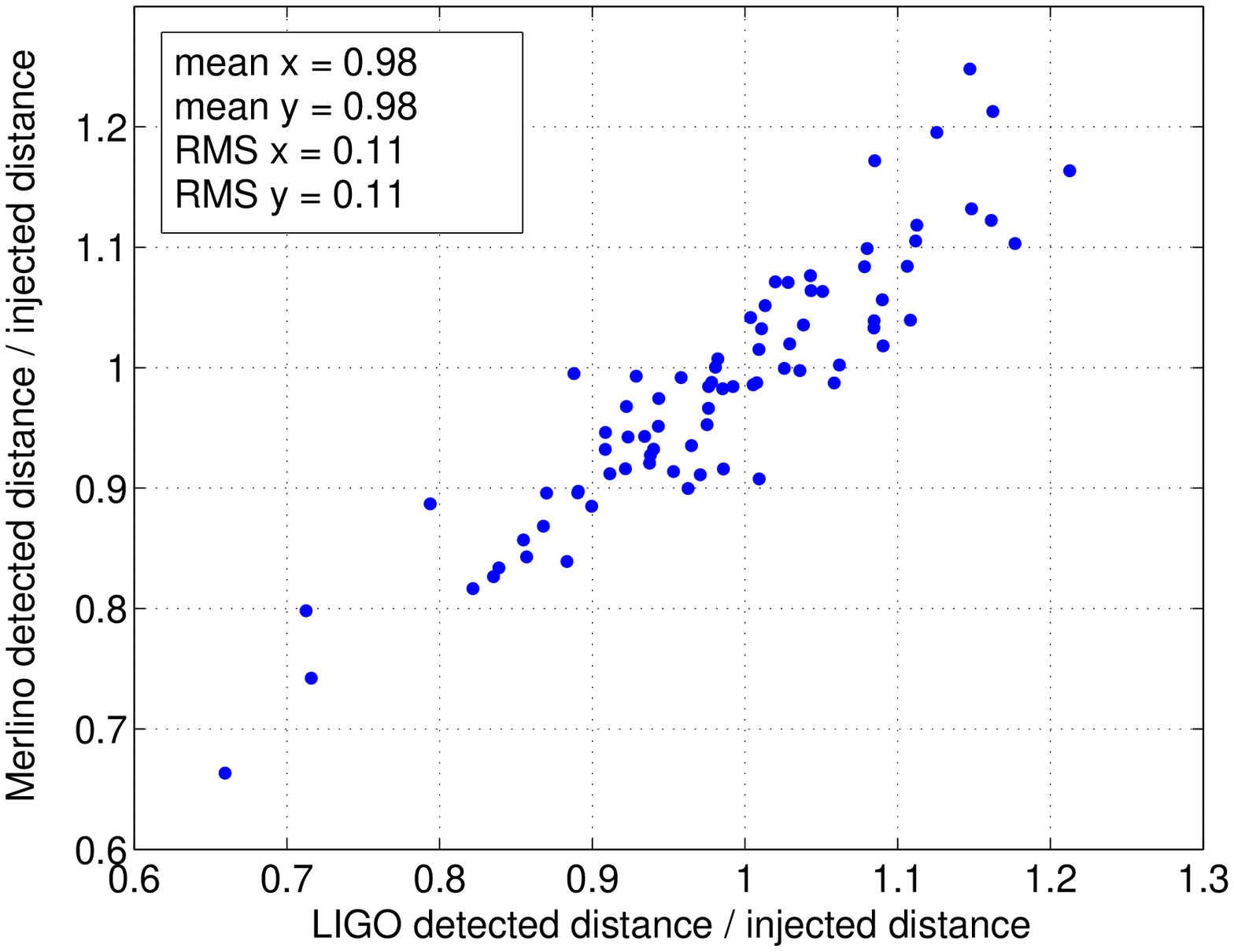}
  \end{center}
\caption{On the left is a scatter plot of the effective distance determination ratio for MBTA and LIGO events detected by both pipelines in the V1 data; on the right is the same but for Merlino and LIGO events. The parameter determination ratio is defined as the detected effective distance divided by the actual injected effective distance.}
\label{compMBTAdist}
\end{figure}

Fig.~\ref{compMBTAchirp} (left) displays a comparison of the accuracy of the determination of the chirp mass for events detected by both the MBTA and LIGO pipelines from within the V1 data set, while a similar chirp mass comparison from the LIGO and Merlino pipelines is displayed in Fig.~\ref{compMBTAchirp} (right). In Fig.~\ref{compMBTAdist} (left) we see the recovered effective distance divided by the actual injected effective distance for the signals detected by MBTA and LIGO in the V1 data, while Fig.\ref{compMBTAdist} (right) shows a similar effective distance comparison between the Merlino and LIGO pipelines. The chirp mass and effective distance estimates by LIGO, Merlino and MBTA were essentially the same.

\section{Benefits from a Combined LIGO-Virgo inspiral search}
\label{benes}
This close examination of the LIGO and Virgo binary inspiral search pipelines has also provided additional information on the benefits one can get in conducting a mutual search for signals. In the effort to detect binary inspiral gravitational waves the inclusion of Virgo data significantly increases the probability for observing a coincident signal in at least two interferometers. The results presented above show that each pipeline is able to effectively detect events with sufficiently low effective distance, and recover the chirp mass, coalescence time, and effective distance. Since the effective distance value is influenced by detector orientation and source location, it is not a parameter that can be used as a test for coincident detection. However, the chirp mass and coalescence time can be used to require consistency. We set coincident window sizes of $0.2 M_\odot$ for the chirp mass and $\pm 8$ ms about the light travel time between interferometers for the coalescence time (for reference, the light travel times between the detectors are 27.2 ms for Hanford - Virgo, 26.4 ms for Livingston - Virgo, and 10 ms for Hanford - Livingston), and a $SNR > 6$ threshold. With these settings we found no triple coincident false events, and only one double coincident false alarm. 

The double coincidence detection ability for the injected signals increased with the inclusion of V1 over just the H1-L1 coincidence alone. This is summarized in Table 1, where the triple coincidence results and various two detector coincidence results are presented. The efficiency for detection of injections from NGC 6744 is larger since it is closer than M87.

\begin{table}[ht]
\begin{center}
\begin{tabular}{c c c c c c}
\hline \hline
 & HLV & HL & HV & LV & HL $\cup$ HV $\cup$ LV \\
\hline
 NGC 6744 efficiency & 48\% & 65\% & 54\% & 49\% & 72\% \\
 M87 efficiency & 24\% & 42\% & 32\% & 30\% & 56\% \\
\hline\hline
\end{tabular} 
\end{center} 
\caption{The efficiency of detecting inspiral injections from NGC 6744
(at a distance of 10 Mpc) and M87 in the Virgo cluster (at a distance of
16 Mpc) using different combinations of the LIGO and Virgo detectors and
an SNR threshold of 6 in all detectors. These results are from the MBTA pipeline.}  
\label{tab:inspiral_coinc_eff} 
\end{table}

The coincidence results show the benefits of performing a search
including all three detectors.  The highest efficiency is obtained by
requiring a signal to be observed in any two of the three detectors.
For the closer NGC 6744 galaxy, the main advantage of adding the Virgo
detector to a LIGO only search is the good triple coincident efficiency.
Not only is the triple coincident false alarm rate very low, but also
with a trigger in three detectors we can reconstruct the sky location of
the source.

For signals from both M87 and NGC 6744, the two detector LIGO efficiency is greater than
either the H1-V1 or L1-V1 efficiency.  This is expected due to the
similar orientations of the two LIGO detectors.  However, by including
Virgo and requiring a coincident trigger in two of the three detectors,
we do obtain a 25\% relative increase in efficiency.  The M87 galaxy is in the
Virgo cluster, which contains a significant fraction of potential binary
neutron star inspiral sources for the initial interferometric detectors.
A 25\% increase in efficiency to these sources significantly increases
the chance of making a detection.

\begin{figure}[tb]
  \begin{center}
    \includegraphics[width=2.6in,angle=0]{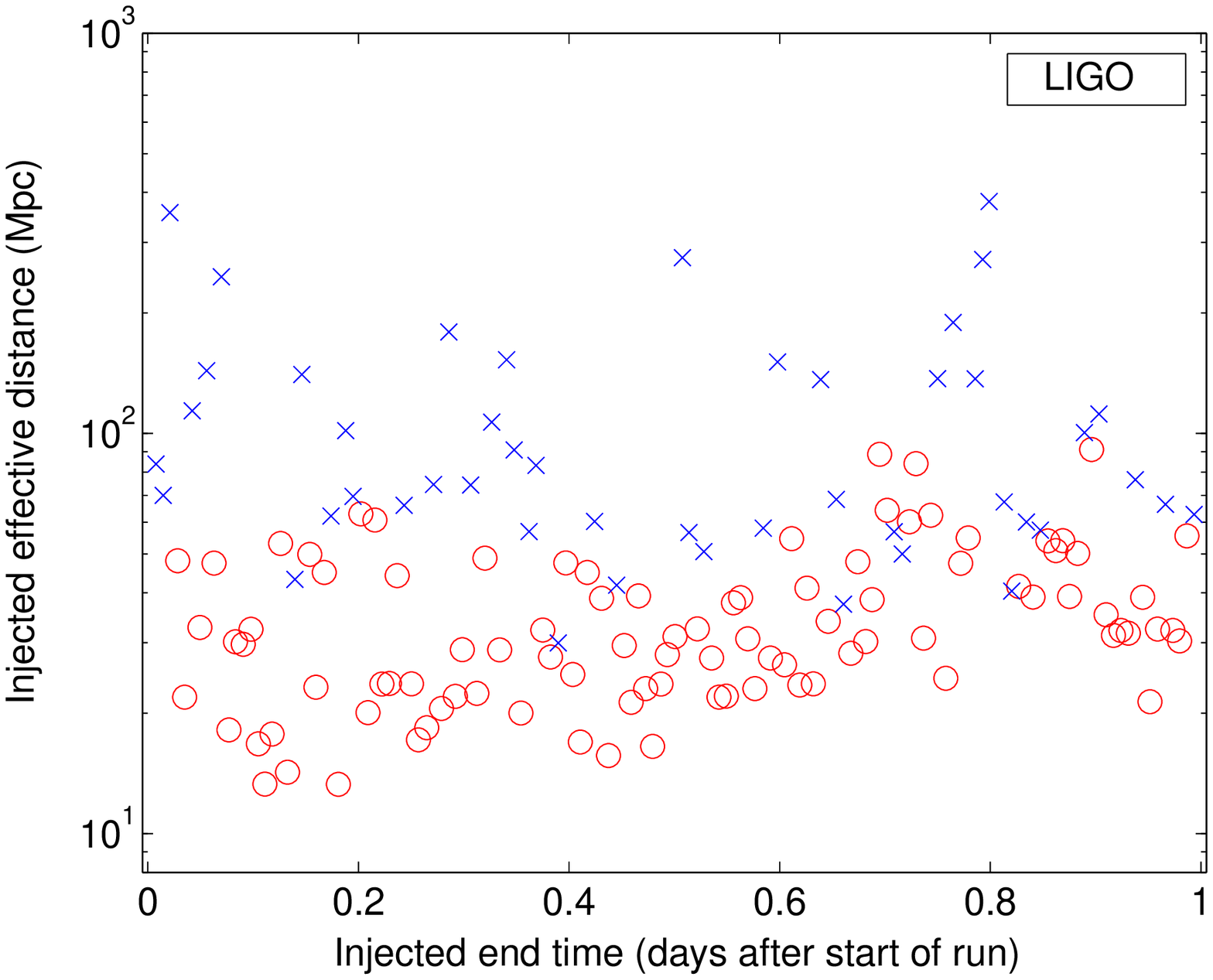}
    \includegraphics[width=2.6in,angle=0]{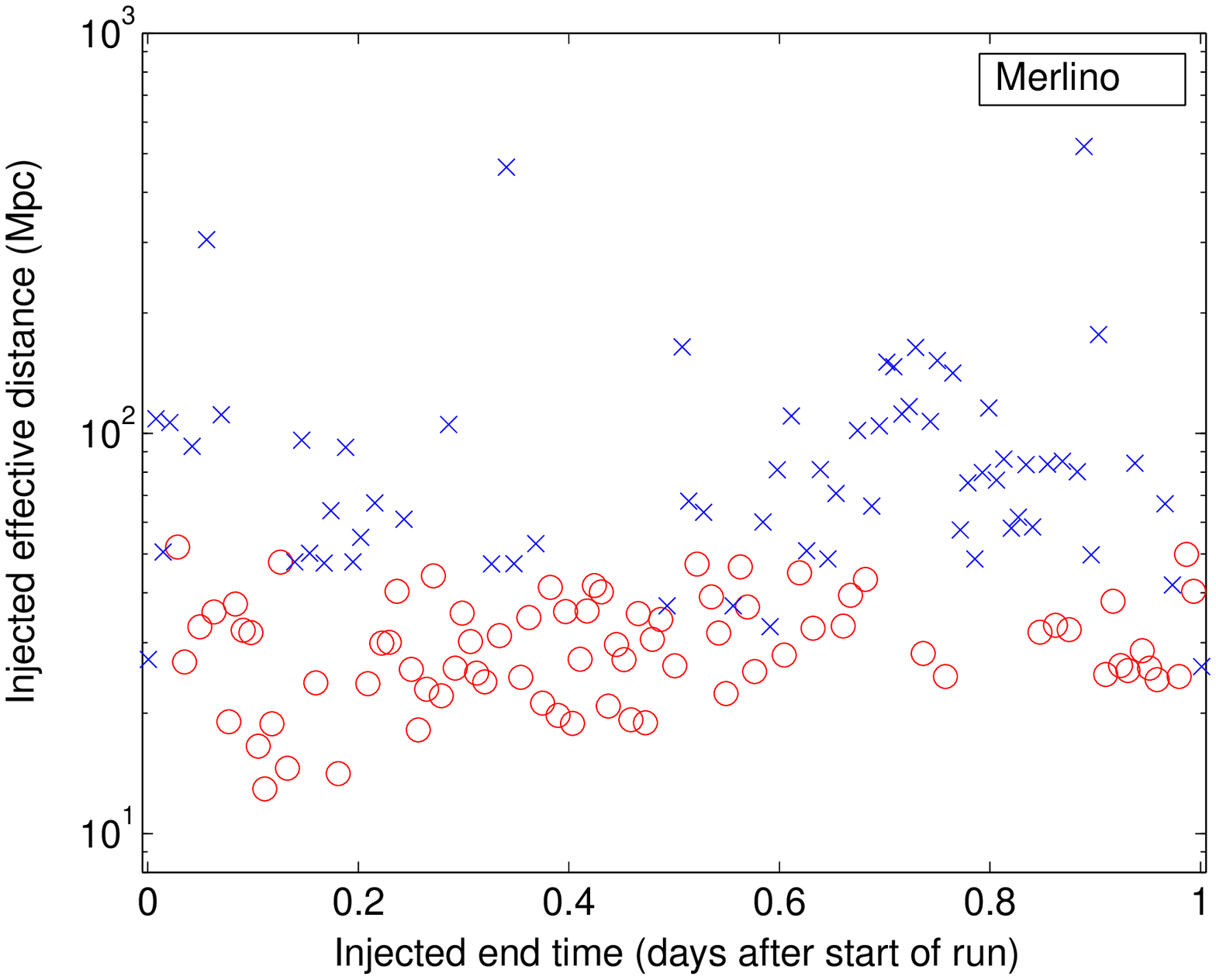}
    \includegraphics[width=2.6in,angle=0]{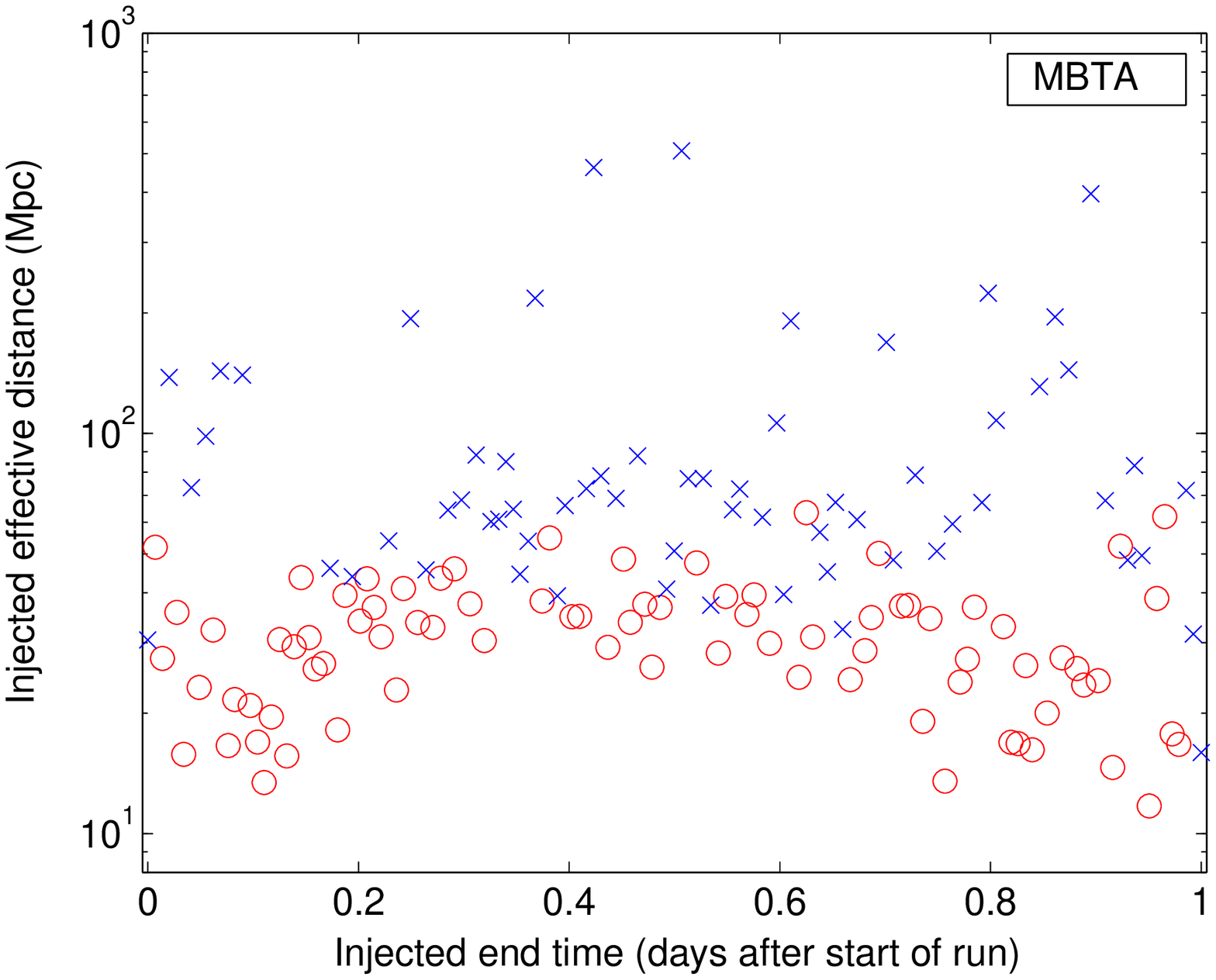}
  \end{center}
\caption{The detected (o) and missed (x) events for the LIGO pipeline on the H1 data, the Virgo Merlino pipeline on the L1 data, and the Virgo MBTA pipeline on the V1 signals. Note too that the dependence of detection efficiency versus effective distance can be seen in these plots; events with an effective distance in excess of 50 Mpc are difficult to detect.}
\label{figMissed}
\end{figure}

The reason for the increase in two detector efficiency can be seen in Fig.~\ref{figMissed}. Displayed are the detected and missed signals for the LIGO pipeline on the H1 data, the Virgo Merlino pipeline on the L1 data, and the Virgo MBTA for the V1 signals. One can see in these plots that the reason for signals being undetected, in all three pipelines, is due to large effective distance values. It can be seen that over the course of 24 sidereal hours there is a variation in the ability to detect signals from the two source galaxies due to changes in interferometer alignment as the earth rotates. Virgo's inclusion provides detections at times when the LIGO interferometers' alignments may be sub-optimal.

\subsection{Source Directional Information}
\label{dir}
\begin{figure}
\includegraphics[width=18pc]{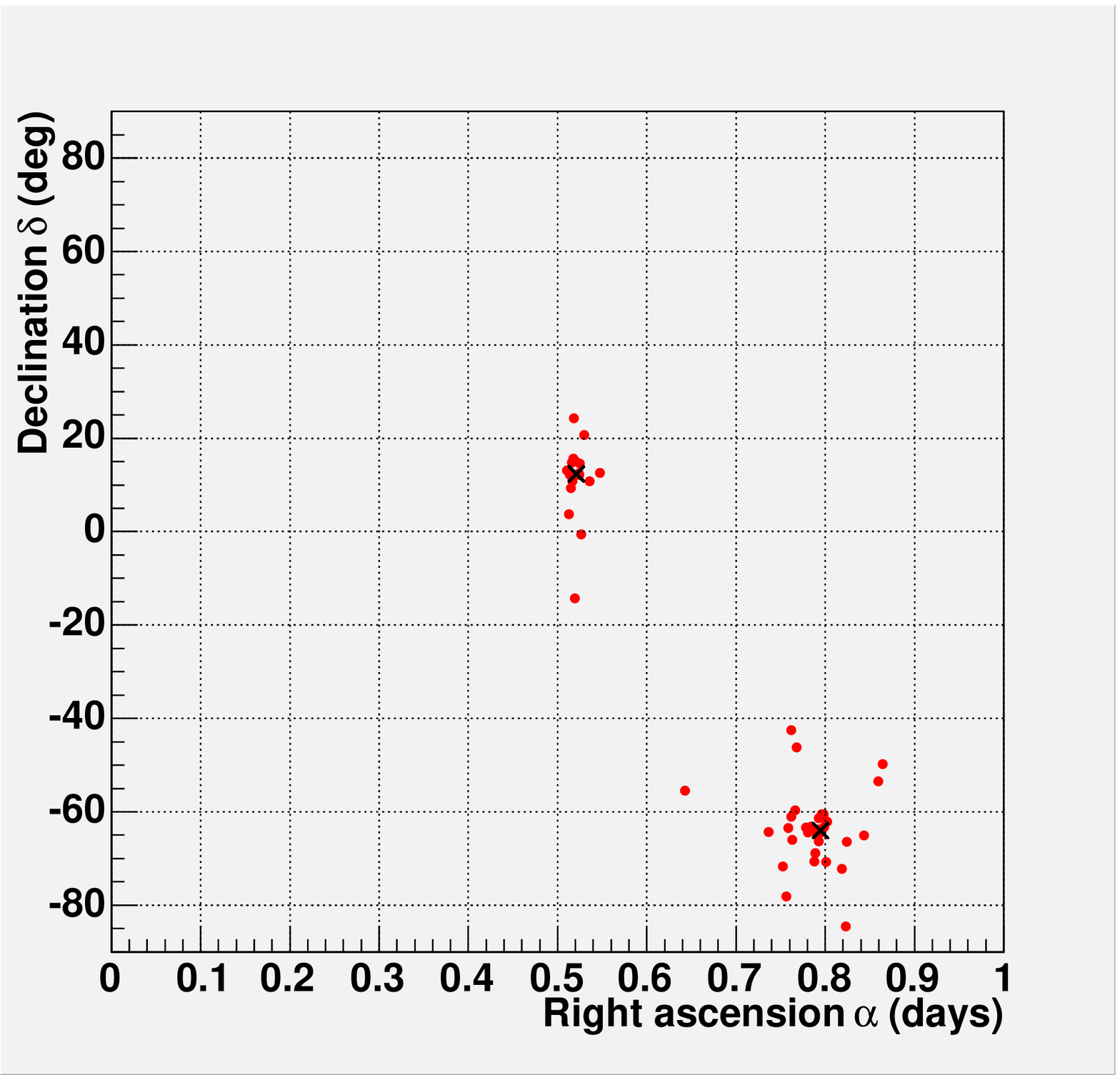}
\begin{minipage}[b]{18pc}
\caption{The recovered (dots) and injected (crosses) sky locations of the inspiral
injections seen in all three detectors.  For reference, the galaxy NGC
is located at $\alpha = 19\ hr\ 9.8\ m$, $\delta = -64^{\circ}$ and M87 is
located at $\alpha = 12\ hr\ 30.8\ m$, $\delta = 12^{\circ}$.}      
\label{fig:inspiral_direction}
\end{minipage}
\end{figure}

The use of three widely spaced interferometers provides the opportunity for identifying the location on the sky of a binary inspiral event. This is another positive outcome of a combined LIGO and Virgo search. Fig.~\ref{fig:inspiral_direction} shows the recovered sky position for events that were successfully detected in H1, L1 and V1. The MBTA pipeline was used to identify the triple coincidences using clustered triggers; for this demonstration we selected the highest SNR trigger within $\pm 10$ ms of injection, in each interferometer.

\begin{figure}
\includegraphics[width=18pc]{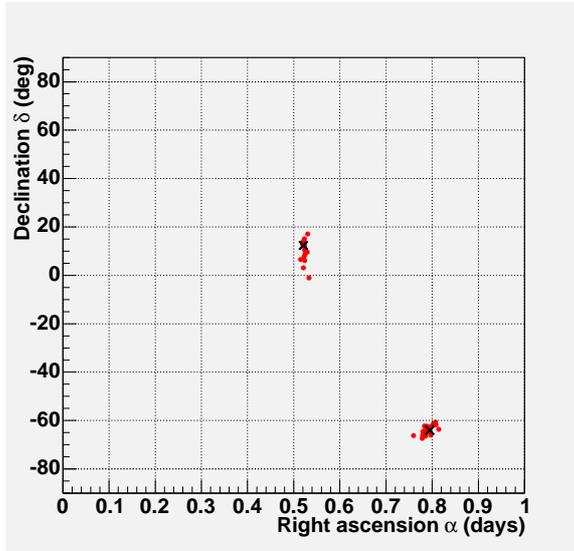}
\begin{minipage}[b]{18pc}
\caption{The recovered and injected sky locations of the subset of inspiral injections which are seen in all three detectors using exactly the same template in $m_1$ $m_2$ space. The 39 detected events in this figure were the ones found to have the same mass parameters for the triggers, as described in the text.}      
\label{Better_inspiral_direction}
\end{minipage}
\end{figure}

The determination of the sky position is affected by the estimates for the other parameters, as the reconstructed values of the parameters are not independent. For example, a higher mass binary inspiral will traverse the sensitive band of the detectors more rapidly than one of lower mass. The reconstructed coalescence time and masses of the system will be correlated.  These correlations make it difficult to determine the coalescence time, and then sky position, with good accuracy. It was possible, however, to improve on the sky position determination. Again using the MBTA pipeline, the search for a signal in H1 and L1 data was restricted such that the identified clusters for individual triggers were only those issued by the same template as the one leading the identified cluster in the V1 data (choosing the highest SNR if more than one is found). Specifically, the analysis of the H1 and L1 data was redone with the same template grid as was used to analyze the V1 data.
However, if no trigger with the same templates as in V1 can be found in H1
or L1, then L1 and V1 are examined for triggers with the same masses as in
H1, and if not successful do the same with L1 as a reference. By conducting the search accordingly 
the ability to estimate the mass parameters (and hence the end time and sky position) improves, and we end up with 39 injections (out of 49 triple coincident detections) found with the same template in $m_1$ $m_2$ space. The order in which each
interferometer is used in turn as a reference is somewhat arbitrary. However for the 
present study the reason
for starting with Virgo is due to the better mass resolution of using the 
V1 data (due to the better low frequency sensitivity). Fig.~\ref{Better_inspiral_direction} shows the sky position accuracy obtained after enforcing this mass correction technique. The position accuracy distribution (recovered position minus true position) for the 16 events from galaxy M87 had a mean of $4.1^o$ and an RMS of $3.5^o$, while for the 23 events from galaxy NGC 6744 the position accuracy distribution had a mean of $2.3^o$ and an RMS of $1.2^o$.

\section{Summary} 
\label{Disc}
LIGO and Virgo both have efficient data analysis pipelines for detecting gravitational waves from the inspiral of binary neutron star systems. It is likely that in the near future LIGO and Virgo will undertake a mutual search for binary inspiral signals. Both groups now have confidence in each other's ability to accurately detect these signals. The results presented in this paper validate the performance of all the LIGO and Virgo inspiral search pipelines.

A combined search for inspiral signals by LIGO and Virgo will also produce significant advantages.  The probability for a event to be seen simultaneously by two detectors increases significantly when Virgo is included with LIGO in an inspiral search. Also, the ability to locate the sky position of an inspiral event becomes possible when observed by Virgo and interferometers at the two LIGO locations. Our immediate future data analysis goals involve moving to the analysis of real data from the LIGO and Virgo interferometers. This will create additional issues that will need to be studied, such as data quality criteria and the use of veto channels.

\acknowledgments
LIGO Laboratory and the LIGO Scientific Collaboration gratefully acknowledge
the support of the United States National Science Foundation for the construction and operation of the LIGO Laboratory and for the support of this research. N.C. also acknowledges the support of the Fulbright Scholar Program. Virgo is supported by the CNRS (France) and INFN (Italy).

\end{document}